\documentclass[twocolumn]{aastex62}
\usepackage{import}

\usepackage{mathtools}
\usepackage{chngcntr}
\usepackage[hang]{footmisc}
\usepackage[caption=false]{subfig}
\usepackage{float}
\usepackage{tabulary,booktabs}
\usepackage{multirow}
\usepackage{graphicx}
\usepackage{amsmath}
\usepackage{yhmath}
\usepackage{mathtools}
\usepackage{csquotes}
\newcommand{\add}[1]{#1}

\usepackage{tikz}
\usepackage{tikz}
\usetikzlibrary{positioning}
\newdimen\nodeDist
\def\Cloudy{\texttt{Cloudy}}
\def\zqual{\texttt{zqual}}

\shorttitle{LATIS MZR}
\shortauthors{Chartab et al.}

\begin{document}

\title{\textbf{LATIS: The Stellar Mass-Metallicity Relation of Star-forming Galaxies at
$z\sim 2.5$}}

\correspondingauthor{Nima Chartab}
\email{nchartab@carnegiescience.edu}

\author[0000-0003-3691-937X]{Nima Chartab}
\affiliation{The Observatories of the Carnegie Institution for Science, 813 Santa Barbara St., Pasadena, CA 91101, USA}

\author[0000-0001-7769-8660]{Andrew B. Newman}
\affiliation{The Observatories of the Carnegie Institution for Science, 813 Santa Barbara St., Pasadena, CA 91101, USA}

\author[0000-0002-8459-5413]{Gwen C. Rudie}
\affiliation{The Observatories of the Carnegie Institution for Science, 813 Santa Barbara St., Pasadena, CA 91101, USA}

\author[0000-0003-4218-3944]{Guillermo A. Blanc}
\affiliation{The Observatories of the Carnegie Institution for Science, 813 Santa Barbara St., Pasadena, CA 91101, USA}
\affiliation{Departamento de Astronom\'{i}a, Universidad de Chile, Camino del Observatorio 1515, Las Condes, Santiago, Chile}

\author[0000-0003-4727-4327]{Daniel D. Kelson}
\affiliation{The Observatories of the Carnegie Institution for Science, 813 Santa Barbara St., Pasadena, CA 91101, USA}

\begin{abstract}
\label{abstract}

We present the stellar mass - stellar metallicity relation for 3491 star-forming galaxies at $2 \lesssim z \lesssim 3$ using rest-frame far-ultraviolet (FUV) spectra from the Ly$\alpha$ Tomography IMACS
Survey (LATIS). We fit stellar population synthesis models from the Binary Population And Spectral Synthesis code (BPASS v$2.2.1$) to
medium resolution (R $\sim 1000$) and  high signal-to-noise ($>30$ per 100 km/s over a
wavelength range of 1221 - 1800 \r{A}) composite spectra of galaxies in bins of stellar mass to determine their stellar metallicity, primarily tracing $\rm Fe/H$. We find a strong correlation between stellar mass and stellar metallicity, with stellar metallicity monotonically increasing with stellar mass at low masses and flattening at high masses ($M_* \gtrsim 10^{10.3} M_\odot$). Additionally, we compare our stellar metallicity measurements with the gas-phase oxygen abundance of galaxies at similar redshift and estimate the average $\rm [\alpha/Fe] \sim 0.6$. Such high $\alpha$-enhancement indicates that high-redshift galaxies have not yet undergone significant iron enrichment through Type Ia supernovae. Moreover, we utilize an analytic chemical evolution model to constrain the mass loading parameter of galactic winds as a function of stellar mass. We find that as the stellar mass increases, the mass loading parameter decreases. The parameter then flattens or reaches a turning point at around $M_* \sim 10^{10.5} M_\odot$. Our findings may signal the onset of black hole-driven outflows at $z \sim 2.5$ for galaxies with $M_* \gtrsim 10^{10.5} M_\odot$.

\end{abstract}

\keywords{\small{Metallicity (1031); Galaxy evolution (594); High-redshift galaxies (734)}}

\section{Introduction}
\label{sec:Introduction}
Cosmic primordial gas is predominantly composed of hydrogen and helium, with small amounts of other light elements such as lithium. The gas accreted from the intergalactic medium (IGM) and circumgalactic medium (CGM) enables a galaxy to form stars, which produce heavy elements. Feedback processes such as stellar winds and supernova explosions expel some of these heavy elements into the interstellar medium (ISM) where new stars are born. Moreover, galactic outflows driven by supernovae and black hole feedback can transfer some enriched material to the IGM and CGM \citep[e.g.,][]{tre04,Chisholm18}. Therefore, the metal content of galaxies is linked to their fundamental evolutionary processes (e.g., star formation and inflow/outflow), and determining its relationship to global properties, such as stellar mass ($M_*$), provides useful information for constraining galaxy evolution models \citep[see review by][]{Maiolino19}. 

The metal content of high redshift galaxies ($2\lesssim z \lesssim 3$) is typically measured in the gas phase (metallicity of ISM) using strong rest-frame optical emission line diagnostics \cite[e.g.,][]{pet04} that are
calibrated locally, suffering significant uncertainties. High-redshift galaxies have been observed to exhibit distinct physical conditions in their H {\sc ii} regions compared to those at $z = 0$ \citep[e.g.,][]{Erb2006,Steidel14,sha2015}, implying a potential evolution of metallicity calibrations with redshift. For direct estimates of gas-phase metallicities, faint auroral lines (e.g., [O{\sc iii}] $\lambda$4363) need to be detected to determine the electron temperature of the ionized gas,  which is now possible for a statistically significant number of high redshift galaxies thanks to the James Webb Space Telescope (e.g., \citealt{san2023}, \citealt{cur2003}, CECILIA survey; \citealt{strom21}, AURORA survey; \citealt{shap2021}). 

Alternatively, stellar continuum emission can be used to measure the metal content of galaxies. Deep optical spectroscopy of $z\sim2-3$ galaxies allows us to access the rest-frame far-ultraviolet (FUV) part of the spectrum that contains important information about their underlying stellar population. Most of the emission in the FUV originates from short-lived O and B stars, and the inferred metallicities are expected to be similar to those derived for the ISM, out of which these young stars have recently formed. The photospheres of hot O and B stars, metal-dependent stellar winds, and interstellar lines all contribute to the FUV absorption features. These features usually have complex dependencies on age, metallicity, and initial mass function (IMF). A number of indices (e.g., 1425 \r{A} and 1978 \r{A} indices) have been identified that are optimized to depend only or mostly on metallicity \citep[e.g.,][]{lei2001,rix2004,som2012}.  

In recent studies, full spectral fitting has been used to measure the stellar metallicity of high-redshift galaxies \citep[e.g.,][]{ste2016,cul2019,kri2019,top2020,kash2022,car2022}. This method has the advantage of using all of the information in the spectra. \cite{ste2016} used a composite rest-frame UV and optical spectrum of 30 star-forming galaxies at $z=2.4$ and found them to be $\alpha$-enhanced relative to the solar abundances by a factor of 4-5. While Type Ia supernovae (SNe Ia) and core-collapse supernovae (CCSNe) both produce iron peak elements, $\alpha$-elements are only produced by massive, short-lived stars that explode as CCSNe. Thus, the $\alpha$/Fe abundance ratio is a powerful tool for constraining the relative contribution of SNe Ia and CCSNe. Based on $\rm [\alpha/Fe]\sim 0.6$, they conclude that CCSNe dominate the enrichment of $z\sim 2$ star-forming galaxies. SNe Ia provide enrichment over long timescales (1-3 Gyr) \citep{mao2012}, while young high redshift galaxies with ages $\lesssim 1$ Gyr are not sufficiently old for iron enrichment \citep{str2017}.   

Although it is now well established that there is a strong positive correlation between the gas-phase metallicity and stellar mass of galaxies out to $z\sim 3.5$ \citep[e.g.,][]{Erb2006,Steidel14,san20,strom2022}, there are only a few studies on the relationship between stellar metallicity and stellar mass, especially at high redshift \citep{cul2019,cal2021,kash2022}. \cite{cul2019} used stacks of 681 star-forming galaxies at $z=$ 2.5–5 from the VANDELS survey \citep{pen2018} with a spectral resolution of $R\sim 580$, and \cite{kash2022} utilized 1336 star-forming galaxies at $z=$ 1.6–3 drawn from zCOSMOS-deep survey ($R\sim 200$) \citep{lil2007} to study the stellar mass - stellar metallicity relation (hereafter stellar MZR) around cosmic noon.  

Recently, the Ly$\alpha$ Tomography IMACS Survey \citep[LATIS;][]{Newman20} has obtained deep optical spectroscopy of $\sim 3800$ star-forming galaxies at $2\lesssim z\lesssim 3$ with a spectral resolution of $R \sim 1000$. Compared to earlier studies of the stellar MZR, this data set is substantially larger and benefits from significantly higher resolution.
 
In this paper, we utilize the Binary Population And Spectral Synthesis code \citep[BPASS v2.2.1;][]{eld2017,stan2018} models to constrain the $z\sim2.5$ stellar MZR by fitting composite rest-frame FUV spectra of 3491 galaxies spanning a wide range of stellar masses, $10^{9}M_\odot\leq M_*\leq 10^{11.5}M_\odot$. We do not employ the latest release of BPASS models \citep[v2.3;][]{byr2022} with $\alpha$-enhanced spectra. Although these $\alpha$-enhanced models are highly desired for applications at high redshift, they are limited to Main Sequence/Giant Branch stars while the spectra of OB and Wolf-Rayet stars remain unchanged from BPASS v2.2.1 and are not $\alpha$-enhanced (private communication, C.~Byrne), making them unsuitable for fitting FUV spectra dominated by OB stars. The paper is organized as follows. In Section \ref{sec:Data}, we present an overview of the LATIS survey and details of the sample used in this work. We then describe our spectral analysis to estimate the stellar metallicities in Sections \ref{sec:smodel} and \ref{sec:fitting}. Our results are presented in Section \ref{sec:Results}. We discuss our results in Section \ref{sec:Discussion} and summarize them in Section \ref{sec:Summary}.

Throughout this work, we assume a flat $\Lambda$CDM cosmology with $H_0=70 \rm \ kms^{-1} Mpc^{-1}$, $\Omega_{m_{0}}=0.3$ and $\Omega_{\Lambda_{0}}=0.7$. All magnitudes are expressed in the AB system, and the physical parameters are measured assuming a \cite{Chabrier03} IMF. We adopt the solar metallicity values of $Z_\odot = 0.0142$ and $\rm 12 +log(O/H)_\odot=8.69$ \citep{asp2009}. In this abundance scale, the solar oxygen and iron mass fractions are 0.00561 and 0.00128, respectively.

\section{Data}
\label{sec:Data}
The LATIS survey is a five-year program (2017-2022) conducted using the Inamori-Magellan Areal Camera and Spectrograph \citep[IMACS;][]{Dre11} at the Magellan Baade telescope. The primary goal of LATIS is producing three-dimensional maps of the $z\sim 2.5$ IGM at Mpc resolution, as traced by Lyman-$\alpha$ absorption. LATIS densely sampled Lyman-break galaxies (LBGs) in three legacy fields, the Cosmic Evolution Survey \citep[COSMOS;][]{sco13} and the Canada–France–Hawaii Telescope Legacy Survey (CFHTLS) D1/D4 fields. The final release catalog of CFHTLS (T0007)\footnote{http://terapix.calet.org/terapix.iap.fr/cplt/T0007/doc/
T0007-doc.html} and the \cite{Ilbert2009} catalog of COSMOS were used for selection of LBGs in the CFHTLS D1/D4 and COSMOS fields, respectively. Galaxies were selected based on either their photometric redshifts or $ugr$ optical colors with $r-$band magnitudes $23.0 < r < 24.8$. The observations were performed using the custom grism and filter installed on IMACS resulting in a spectral resolving power of $R = 990$ and a spectral coverage of 3890-5830 \r{A}. This spectral resolution is estimated at the midpoint of the wavelength coverage and is derived from the average size of targets, which are not point-like, in the two-dimensional spectra. For a full description of the survey strategy, observation and data reduction, we refer readers to \cite{Newman20}. 

In the present paper, we use a sample of $z>1.5$ galaxies with high-confidence spectroscopic redshifts based on multiple lines and a well-modeled spectrum (i.e., \zqual{=3} or \zqual{=4} defined in \citealt{Newman20}). \add{
Out of the 7408 galaxies that were initially targeted with LATIS, 6568 received deep exposures as part of the main survey while 840 were observed on backup masks of bright candidates intended for poor conditions. A total of 5443 received a redshift quality flag ($\zqual$) of 3 or 4 and had no severe data reduction problems. After further exclusion of 45 AGNs, 105 QSOs, and 44 spectra comprised of two heavily blended sources, we are left with 5249 galaxies, of which 3888 have a redshift $z>1.5$. In the following sections we refine our sample further by applying stellar mass limits and requiring near-IR photometry, which narrows our sample to 3491 galaxies with \( \rm 10^9M_\odot\leq M_*\leq 10^{11.5}M_\odot \).
}  

\subsection{Photometry}
\label{sec:photometry}

Multi-wavelength coverage from the UV to mid-infrared is essential for fitting the spectral energy distributions (SEDs) of our sample and measuring physical parameters, such as stellar mass and star formation rate (SFR) \citep[e.g.,][]{char2023}{}{}. The photometry of the LATIS galaxies in the COSMOS field is taken from the latest version of the COSMOS catalog \citep[COSMOS2020;][]{Weaver21}. In the CFHTLS D1 and D4 fields, we construct new photometric catalogs by combining optical to near-infrared (NIR) photometry from multiple surveys. In both fields, we use $ugriz$ images from the CFHTLS release T0007 and Spitzer/IRAC channel 1 and 2 mosaics from \cite{ann2018}. In the D1 field, we use $zYJHK_s$ images from the VIDEO survey \citep{jar2013}, while in the D4 field we use $JHK_s$ images from the WIRDS survey \citep{bie2012}. Images are registered and point spread function (PSF)-matched following the procedures described by \cite{new2012}. We use {\tt SWarp} \citep{ber2010} to construct a $\chi^2$ detection image from the NIR (VIDEO or WIRDS) images available in each field, and run {\tt SExtractor} \citep{ber1996} in dual-image model to construct catalogs from the PSF-matched images. We measure colors in $2''$ diameter apertures and scale the fluxes to match {\tt FLUX\_AUTO} in the $r$ band. For the distinct treatment of the IRAC bands, see \cite{new2012}.

When using the zeropoints provided by each survey, we find that the optical-NIR and NIR-IRAC colors of D1/D4 LATIS galaxies are inconsistent with those in the COSMOS field. This seems to arise, at least in part, from a flux-dependent PSF in the NIR images of the D1 and D4 fields; this effect means that matching the PSFs measured with stars does not match the PSFs applicable to faint galaxies. We therefore adjust the NIR and IRAC zeropoints to match the median optical-NIR and NIR-IRAC colors of faint galaxies to those measured in the COSMOS2015 catalog \citep[][using the COSMOS2020 catalog instead results in insignificant differences]{lai2015}. These zeropoint corrections are usually $\approx 0.1$ mag and are fairly insensitive to the flux limits. After applying this procedure, we find that the average SEDs of the LATIS galaxies, as well as the derived distributions of stellar masses and SFRs (Section~\ref{sec:SEDfitting}), are quite consistent across all fields.

\begin{figure}
    \centering
    \includegraphics[width=\linewidth]{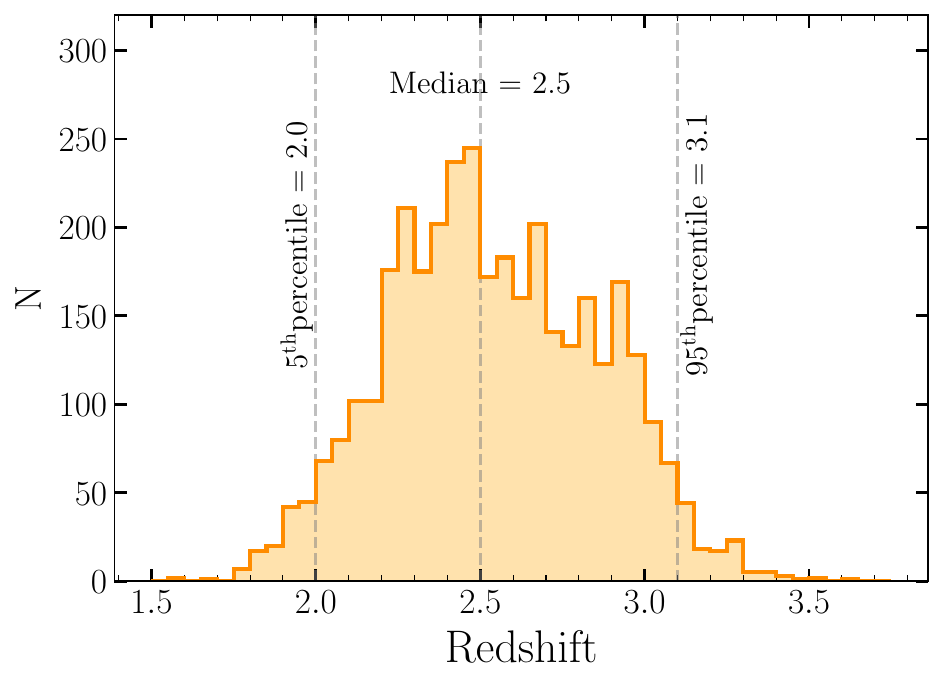}
    \caption{Spectroscopic redshift distribution for the entire  LATIS sample. The vertical dashed lines indicate the median value along with the 5$^{\rm th}$ and 95$^{\rm th}$ percentiles.}
    \label{fig:z}
\end{figure}

We find that 93-97\% of LATIS galaxies in the COSMOS and D1 fields have clear matches in the COSMOS2020 and our newly compiled (optical to NIR) D1 field catalogs. These matches are identified based on the criteria of having separations $< 0\farcs5$ and $\Delta r < 0.4$~mag. The fraction is lower (60\%) in the D4 field, because the WIRDS imaging does not cover the entire area. To ensure that the stellar mass measurements are robust, we require NIR coverage, which  is particularly important in D1/D4 fields where the IRAC data are not as deep. Based on these criteria, we are left with approximately 3500 galaxies with 90\% of the sample at $2.0 \leq z\leq 3.1$ (Figure \ref{fig:z}).

\subsection{Stellar Mass \& Star Formation Rates}\label{sec:SEDfitting}
We fit the UV to mid-infrared SED of galaxies to derive their physical parameters. We use the C++ version of the {\tt LePhare} code \citep{Arnouts,Ilbert} combined with a library of synthetic spectra generated by the \citet{BC03} population synthesis code. The redshifts are fixed to the spectroscopic redshifts from LATIS. Similar to the configuration employed in the COSMOS catalogs \citep{lai2015,Weaver21}, the models incorporate exponentially declining star formation histories with nine $e$-folding times in the range of $0.01<\tau<30\ \text{Gyr}$ and two delayed exponentially declining models (SFR $\propto te^{-t/\tau}$) with $\tau=3$ and $5$ Gyrs. The delayed models are included since high-redshift galaxies likely exhibit star formation histories that differ substantially from simple exponential decays \citep[e.g.,][]{red2012,kel2010}. We adopt the \citet{Chabrier03} IMF, truncated at 0.1 and 100 $ \text{M}_{\odot}$, and the \citet{Calzetti} attenuation law to apply dust extinction ($\rm E(B-V)\leq 1.1$). The code also includes emission lines using the \citet{Kennicutt} relation between SFR and UV luminosity, as described in \citet{Ilbert2009}. Three different stellar metallicities are considered: $Z_*=0.02, 0.008$, and 0.004.

\begin{figure}
    \centering
    \includegraphics[width=\linewidth]{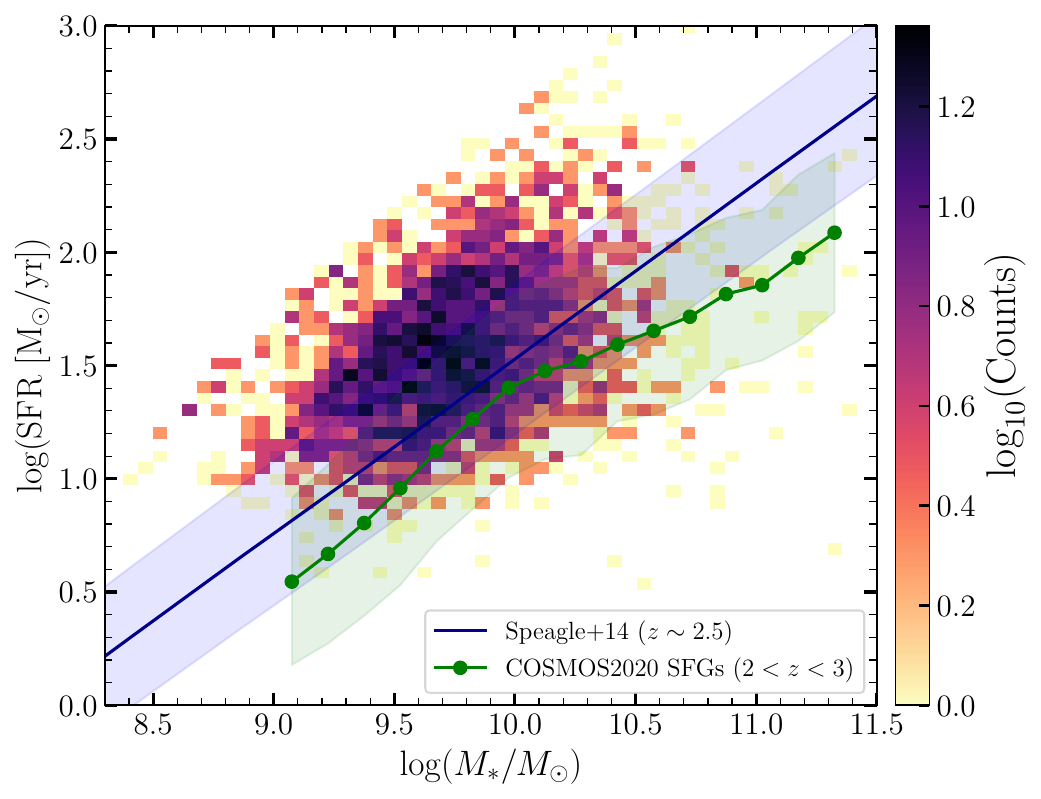}
    \caption{SED-derived SFR as a function of stellar mass for the entire LATIS sample. For comparison, the main-sequence relation from \cite{Speagle14} is also shown. \add{The green line represents the median trend for the star-forming galaxies at $z=2-3$ in the COSMOS2020 catalog.}}
    \label{fig:SFR-M}
\end{figure}

For each template, {\tt LePhare} computes fluxes in all bands, then determines the template with a minimum $\chi^2$ based on the model and observed fluxes. It also provides percentiles of the marginalized posterior probability distributions (probability $\rm \propto e^{-\chi^2/2}$). In this work, we use the median posterior values for stellar mass, SFR, and sSFR (SFR/$M_*$). 


Figure \ref{fig:SFR-M} shows the distribution of SED-derived stellar masses and SFRs. The main sequence (MS) relationship is also shown from \cite{Speagle14} for comparison. Overall our sample is in agreement with the $z \sim 2.5$ MS, although its average SFRs are slightly higher\add{. We further compare to galaxies in the COSMOS2020 catalog, whose masses and star-formation rates were derived using similar techniques, thereby minimizing systematics. The green line in Figure \ref{fig:SFR-M} represents the median trend for the star-forming galaxies selected by $\text{sSFR}>10^{-10.1} \text{yr}^{-1}$ \citep{Pacifici16} and $z=2-3$. At fixed stellar mass, our sample exhibits a $\sim0.3$ dex average enhancement in SFR compared to these galaxies. As a result, our sample predominantly represents star-forming galaxies located on and above the main sequence at cosmic noon.}

\subsection{Composite spectra}
\label{sec:stack_spectra}
\begin{figure*}
    \centering
    \includegraphics[width=\linewidth]{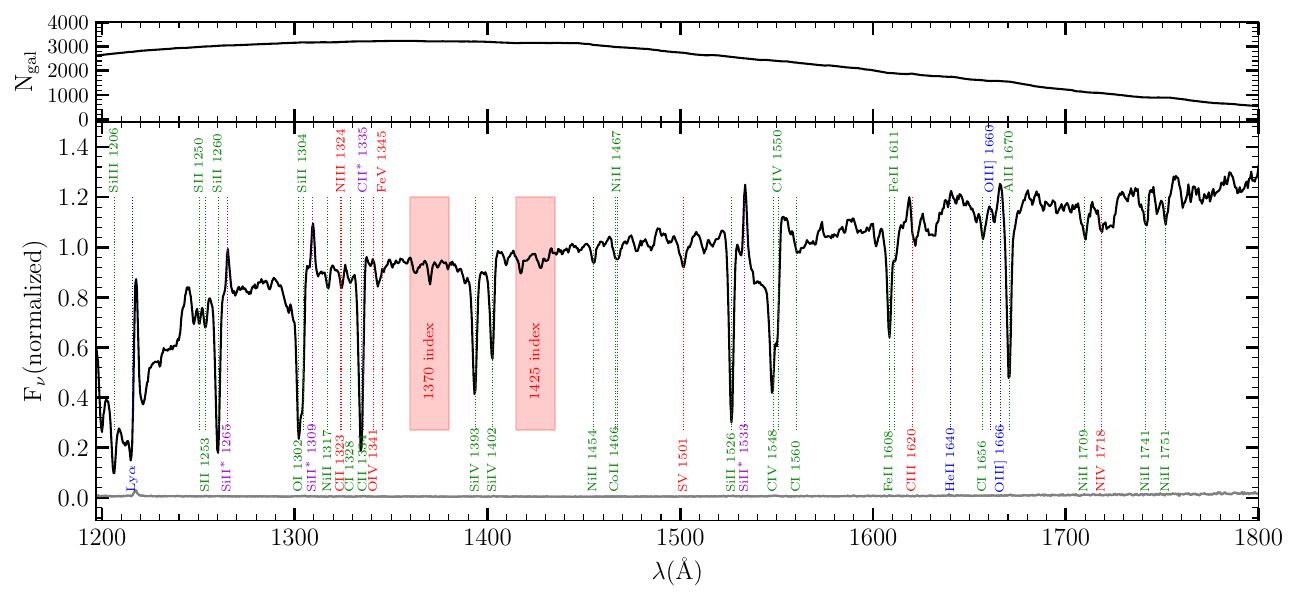}
    \caption{Composite rest-frame FUV spectrum of the entire LATIS sample. The top panel shows the number of galaxies contributing to each pixel in the composite spectrum, ranging from 536 to 3235. The composite spectrum is shown in the bottom panel; some prominent emission and absorption lines are labeled. Stellar and interstellar absorption features are color-coded with red and green, respectively. Several nebular emission lines (blue), and fine structure emission lines (purple) are also included. The red shaded regions indicate wavelength ranges for 1370 and 1425 indices as defined in \cite{lei2001}; these indices cover several photospheric lines such as Fe {\sc v} $\lambda\lambda 1360-1380$, O {\sc v} $\lambda 1371$, Si {\sc iii} $\lambda 1417$,
C {\sc iii} $\lambda 1427$, and Fe {\sc v} $\lambda 1430$, which are not labeled for legibility. The gray line shows the 1$\sigma$ error spectrum.}
    \label{fig:stacked_spectra_full}
\end{figure*}

Most of the prominent absorption and emission features in the FUV are interstellar in origin. Typically, stellar features in the FUV spectrum are weak and require a high signal-to-noise ratio to be useful for measuring stellar metallicity. Gravitationally lensed systems often provide an adequate signal-to-noise ratio for individual galaxies. However, in our case, where individual galaxies lack sufficient signal-to-noise ratio, stacking techniques can be used to boost the signal strength. To construct a composite spectrum, the individual spectra are first corrected for Galactic extinction using the \cite{Schlafly11} dust map, followed by masking pixels with strong skyline contamination. The individual spectra are shifted to the rest frame using their spectroscopic redshifts and then normalized by their median fluxes within 1425-1500 \r{A} to prevent the dominance of high-SFR galaxies in the composite spectrum. The normalized spectra are resampled onto a grid of wavelengths corresponding to $\Delta v=100$ km/s. This results in a wavelength spacing of $\sim 0.4-0.6$ \r{A} in a rest-frame FUV spectrum covering $\sim 1200-1800$ \r{A}. The composite spectrum is determined by the median of the normalized spectra, and the $1\sigma$ error is derived by bootstrapping. Figure \ref{fig:stacked_spectra_full} illustrates the high signal-to-noise composite spectrum of all 3491 galaxies with $\rm 10^9M_\odot\leq M_*\leq 10^{11.5}M_\odot$ included in the present work (S/N=227 per 100 km/s at $\lambda_{\rm rest}\sim$ 1450 \r{A}).           

\subsubsection{Stellar mass bins}
\label{subsec:m_bin}

To study the stellar MZR, we construct composite spectra within bins of stellar mass. The median uncertainty in stellar mass measurements is $\sim 0.15$ dex, which provides a lower limit on the size of the stellar mass bins. Furthermore, we require that a composite spectrum has a median signal-to-noise ratio per 100 $\rm km/s$ of at least 30 over a wavelength range of 1221-1800 \r{A} to ensure a reliable metallicity measurement. Over a similar wavelength range, we require that at least 25 galaxies contribute to every pixel of a composite spectrum in order to obtain reliable $1\sigma$ error estimates. These criteria leave us with nine stellar mass bins, which will be utilized to construct the stellar MZR in Section \ref{sec:stellarmzr}.            
\section{Stellar population synthesis models}
\label{sec:smodel}
To determine the stellar metallicity of galaxies, we compare the observed composite spectra with synthetic stellar population models. We use the models from the Binary Population And Spectral Synthesis code \citep[BPASS v2.2.1;][]{eld2017,stan2018} adopting the \cite{Chabrier03} IMF with a high-mass cutoff of 100 $M_\odot$. 
 The set of simple binary stellar population spectra (i.e., instantaneous starbursts) with an initially formed stellar mass of $10^6$ $M_\odot$ at ages $\rm log(a'/years)=6.0-11$ in $0.1$ dex increments is provided at a pixel resolution of 1~\r{A} for 13 different stellar metallicities ($10^{-5}\leq Z_* \leq 0.040$).

We add nebular continuum to the stellar population models ({\tt BPASS+Nebular}) using the photoionization code \Cloudy{} v17.02 \citep{fer2017}. We model H{\sc ii} regions with a spherical shell geometry of a fixed radius, adopting the BPASS spectrum as the incident spectrum. The metallicity of ionized gas is assumed to be $Z_{\rm neb}=0.5 Z_\odot$, which is the average gas-phase metallicity of galaxies at the redshift interest of our study \citep[e.g.,][]{ste2016,str2018,san2021}. Additionally, we use an ionization parameter of $\log U=-2.8$ and electron
density of $n_e=250\ \rm cm^{-3}$ that are consistent with recent estimates for $z\sim 2-3$ star-forming galaxies \citep[e.g., ][]{san2016b,str2017,top2020}. Due to the minimal contribution of the nebular continuum to the total flux ($\lesssim 10\%$), our results are not sensitive to the assumptions of the \Cloudy{} model.    
\begin{figure*}
    \centering
    \includegraphics[width=\linewidth]{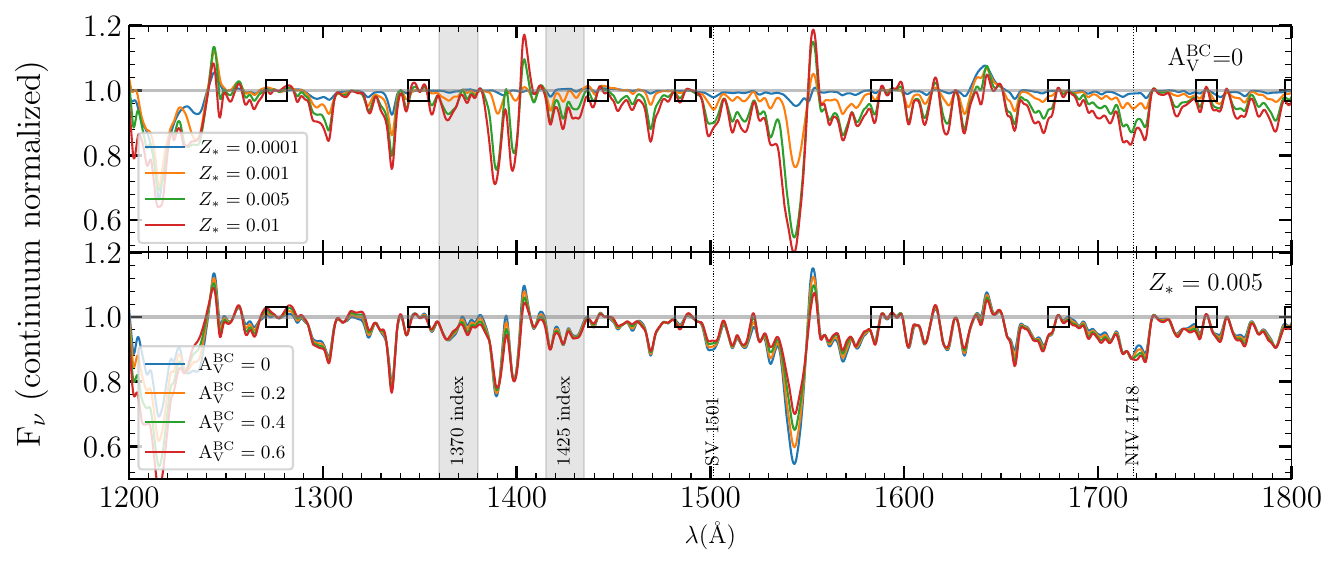}
    \caption{The BPASS v2.2.1 stellar population models at the average effective resolution of the LATIS composite spectra ($\sigma_{\rm v}=250$ km/s) assuming a constant SFR over 100 Myr. The top panel shows the variation of the model with metallicity at a fixed $\rm A_{V}^{BC}=0$. Various metallicities, ranging from $10^{-4}$ to $0.01$, are shown with different colors, as indicated in the legend. The bottom panel shows the sensitivity of the model to $\rm A_{V}^{BC}$ at $Z_*=0.005$. For illustration purposes, we normalized all model spectra using the pseudo-continuum derived from the spline fit to the wavelength windows (black squares) suggested by \cite{rix2004}. }
    \label{fig:BPASS_model}
\end{figure*}

We construct models assuming a constant star formation history by integrating burst models over their ages. The choice of constant star formation history is motivated by the fact that the FUV spectrum is mainly reliant on the past $100\ \rm Myr$ history of star formation, and the star formation history is expected to be relatively constant over this timescale for an ensemble of galaxies at $z\sim 2-3$ \citep[e.g.,][]{red2012,cul2019}. Therefore, model spectra, ${\rm F}(\lambda,a,Z_*)$, are constructed using the following equation: 

\begin{equation}
    {\rm F}=\sum_{i=1}^{{\rm N}} f^{{\rm BPASS}}(\lambda,a'_{i},Z_*)T(\lambda,a'_{i})\Delta a'_{i}
\end{equation}

\noindent where {\tt BPASS+Nebular} models ($f^{{\rm BPASS}}$) are summed up over the age bins ($a'$) such that $\rm N$ is the number of BPASS age bins within the age of a galaxy ($a$), $\Delta a'_{i}$ are the widths of the age bins, and $T(\lambda,a'_{i})$ represents the transmission function due to dust attenuation. A population of stars younger than the lifetime of the stellar birth clouds suffers from greater attenuation than those outside the region \citep{char2000}. The typical lifetime of the birth clouds is 10 Myr \citep[e.g.,][]{bli1980}. We therefore model the transmission function as follows \citep[see also][]{car2022}:

\begin{equation}
     T(\lambda,a'_{i})= T^{\rm ISM}(\lambda)\times\begin{cases}
     T^{\rm BC}(\lambda)      &\ \  \text{if } a'_{i}\leq 10 {\rm\ Myr}\\
    1                  &\ \  \text{otherwise}
\end{cases}
\end{equation} where $T^{\rm ISM}(\lambda)$ and $T^{\rm BC}(\lambda)$ are the transmission functions of the ambient ISM and birth clouds, respectively. In this paper, we do not intend to infer ISM dust attenuation parameters; they are merely nuisance parameters in our model and are coupled with possible flux calibration errors. To properly model any imperfections in spectroscopic calibration, a fifth-order polynomial function is used for the $T^{\rm ISM}(\lambda)$ component; however, the $T^{\rm BC}(\lambda)$ component is modeled using the \cite{Calzetti} dust attenuation, wherein the visual band attenuation $A^{\rm BC}_{\rm V}$ is included as a free parameter. Figure \ref{fig:BPASS_model} shows examples of the model spectra as the metallicity ($Z_*$) and A$^{\rm BC}_{\rm V}$ are varied while the other parameters are fixed. It is evident that absorption features are much stronger as the stellar metallicity increases and that this has the largest effect on the spectrum. Moreover, the figure shows that although the FUV stellar absorption lines are relatively insensitive to the birth-cloud attenuation A$^{\rm BC}_{\rm V}$, the small variation in equivalent width of several features, including the 1718 \r{A} and 1501 \r{A} regions, can be used to capture any covariance between A$^{\rm BC}_{\rm V}$ and $Z_*$, given a high signal-to-noise spectrum.

\section{Fitting Synthesis Model Spectra}
\label{sec:fitting}

In this section, we present our Bayesian full spectrum fitting method, along with the associated parameters and priors. In order to accurately compare our model spectra with the observed composite spectra, it is important to ensure that the spectral resolution is consistent between the two. We first interpolate the {\tt BPASS+Nebular} models to bins of 100 km/s, aligning them with the wavelength grid of the observed spectra. We then introduce a parameter, $\sigma_{\rm v}$, that accounts for the combined effects of galaxies' stellar velocity dispersion, the spectrograph resolution, and redshift uncertainties. We use estimates of these factors to place an informed prior on $\sigma_{\rm v}$ while allowing for uncertainties in these estimates (In Appendix \ref{app:mzr_const_sigma} we demonstrate the effect of adopting a fixed best estimate of $\sigma_{\rm v}$ in our fitting). We convolve our model spectra with a Gaussian kernel in velocity space, resulting in model templates that we refer to as ${{\rm {\tilde{F}}}(\lambda},Z_*,a,\sigma_{\rm v},A^{\rm BC}_{\rm V})$.  

\begin{table}[!h]
\caption{Parameters and priors used for the model fitting}
\label{table:param}
    \centering
    \begin{tabular}{ccc}
    \hline\hline
         Free parameter&  Prior  & Limits \\ \hline
         $Z_*$ & Logarithmic & $(10^{-4},0.04)$ \\
         $a\ ({\rm year})$  & Logarithmic & $(10^7,10^{9.6})$ \\
         $\sigma_{\rm v}$ ($\rm km/s$) & Gaussian ($\mu=165, \sigma=35$) & (100,400) \\
         $\rm A_{\rm V}^{BC}$  & Uniform & (0,4) \\\hline\hline
               
    \end{tabular}
    \label{tab:my_label}
\end{table}

\begin{table*}[!t]

\caption{Rest-wavelength ranges excluded from fitting}
\label{table:mask}
    \centering
    \begin{tabular*}{0.8\textwidth}{@{\extracolsep{\fill}}ccc@{}}

    \hline\hline
         $\rm \lambda_{min}$ (\r{A})&$\rm \lambda_{max}$ (\r{A}) & Interstellar spectral
         features \\ \hline
         1248 &1270 & S {\sc ii} $\lambda$$\lambda$1250.58,1253.81, Si {\sc ii} $\lambda$1260.42, Si {\sc ii}$^*$ $\lambda$1265.00 \\
         1291 & 1320  & O {\sc i} $\lambda$1302.17, Si {\sc ii} $\lambda$1304.37, Si {\sc ii}$^*$ $\lambda$1309.28, Ni {\sc ii} $\lambda$1317.22  \\
         1326 & 1340  &  C {\sc i} $\lambda$1328.83, C {\sc ii} $\lambda$1334.53, C {\sc ii}$^*$ $\lambda$1335.71\\
         1386&  1406 & Si {\sc iv} $\lambda\lambda 1393.76,1402.77$  \\
         1450&  1471 & Ni {\sc ii} $\lambda\lambda\lambda 1454.84,1467.26,1467.76$, Co {\sc ii} $\lambda 1466.21$ \\
     1521 & 1529  & Si {\sc ii} $\lambda 1526.71$ \\
        1531 & 1540  & Si {\sc ii}$^* \lambda 1533.43$  \\
         1543& 1565  & C {\sc iv} $\lambda\lambda$1548.19,1550.77, C {\sc i} $\lambda$1560.31  \\
         1605& 1615  & Fe {\sc ii} $\lambda$1608.45, Fe {\sc ii} $\lambda$1611.20 \\
         1654& 1677  & C {\sc i} $\lambda$1656.93, O {\sc iii]} $\lambda\lambda$1660.81,1666.15, Al {\sc ii} $\lambda$1670.79 \\
        1706 &  1715 & Ni {\sc ii} $\lambda$1709.60  \\
         1737&1755   &  Ni {\sc ii} $\lambda\lambda$1741.55,1751.91\\ \hline\hline
    \end{tabular*}
\end{table*}

We fit the {\tt BPASS+Nebular} models ($\rm \tilde{F}$) to composite spectra within the framework of Bayesian inference. In the case of our problem, the parameter vector of the models is $\boldsymbol{\theta}=(Z_*,a,\sigma_{\rm v},A^{\rm BC}_{\rm V})$, and assuming Gaussian and independent uncertainties in the composite spectra, the logarithm of the likelihood function can be constructed as,

\begin{equation}
    \log\mathcal{L}=-\frac{1}{2}\sum_\lambda \left( \frac{[f(\lambda)-{\rm \tilde{F}}(\lambda,\boldsymbol{\theta})]^2}{\sigma(\lambda)^2}+\ln[2\pi \sigma(\lambda)^2]\right)
\end{equation} where $f(\lambda)$ and $\sigma(\lambda)$ are the flux and 1$\sigma$ uncertainty of the observed composite spectrum, respectively. Due to the fact that our models are provided for discrete values of the parameters, we linearly interpolate ${\rm {\tilde{F}}}$ models across different grids in order to define the likelihood at any given parameter value. Rather than treating $T^{\rm ISM}$ as a free parameter in the model, we determine its value by fitting a fifth-order polynomial function to $f(\lambda)/{\rm {\tilde{F}(\lambda,\boldsymbol{\theta})}}$. This step is performed at every evaluation of the likelihood function during the Bayesian sampling process. This approach ensures that the model's continuum shape matches that of the observed spectra while keeping the computational complexity manageable. $A^{\rm BC}_{\rm V}$ are given a uniform prior, but a logarithmic prior is applied for age and stellar metallicity. A Gaussian prior is adopted for $\sigma_{\rm v}$.
In order to calculate this prior, we consider several factors: the spectral resolution of $\sigma_{\rm inst} = 125\pm10$ km/s achieved for typical galaxies (representing the average and range over the rest-frame wavelength range of 1221-1800 \r{A}, Section~\ref{sec:Data}), random errors in redshifts of $\sigma_z = 93\pm27$ km/s (Newman et al. in prep), and a typical stellar velocity dispersion of $\sigma_* = 100\pm50$ km/s \citep[e.g.,][]{bar2014}. The BPASS models are distributed in wavelength bins of 1~\AA, coarser than our pixels. Binning can be regarded as convolution by a top-hat kernel, which we approximate as a Gaussian with equal FWHM, leading to a model resolution of $\sigma_{\rm mod} = 86 \pm 10$ km/s. Since $\sigma_{\rm v}$ represents the kernel required to match the BPASS models to the observations, it can be computed as $\sigma_{\rm v}^2 = \sigma_{\rm inst}^2 + \sigma_z^2 + \sigma_*^2 - \sigma_{\rm mod}^2$. Combining our estimates of each term results in an effective resolution of $\sigma_{\rm v} = 165\pm 35$ km/s. Table \ref{table:param} presents the list of the parameters and priors. In order to obtain the posterior distribution of the parameters, we use the {\tt MULTINEST} nested sampling algorithm \citep{skil2006,Fer2008,Fer2009,Fer2013,Fer2019}, which can be accessed via the {\tt PYMULTINEST} interface \citep{Buc2014}.

As shown in Figure \ref{fig:stacked_spectra_full}, not every part of FUV spectra originates from stellar emission. Most of the prominent lines in a rest-UV spectrum are predominantly formed in the ISM and/or have a nebular origin. These lines are not included in the model spectra (${\rm {\tilde{F}}}$).  We therefore exclude the wavelength regions contaminated by interstellar absorption or nebular emission lines. The He {\sc ii} $\lambda$1640 emission line is not excluded since this line originates primarily from hot stars that are well incorporated in BPASS models.  Table \ref{table:mask} summarizes the masked regions ignored during the fitting. These masked wavelengths are also shown in Figure \ref{fig:stacked_spectra_full} as shaded regions. We note that our full spectral fitting procedure includes the 1370 and 1425 indices \citep{lei2001} which are attributed to Fe {\sc v} $\lambda\lambda 1360-1380$ and O {\sc v} $\lambda 1371$, and to Si {\sc iii} $\lambda 1417$, C {\sc iii} $\lambda 1427$, and Fe {\sc v} $\lambda 1430$, respectively. The red lines in Figure \ref{fig:stacked_spectra_full} show some other stellar photosphere absorption lines within the wavelength range we considered for fitting ($\rm 1221-1800$ \r{A}). 

When conducting spectral fitting, we employ two iterations to address any potential discrepancies between the observed and model spectrum. During the first iteration, we exclusively mask out the regions where the ISM absorption lines are present, as specified in Table \ref{table:mask}. Following the initial fit, we mask the pixels displaying 3$\sigma$ deviations from the best-fit model and repeat the analysis. We find only a marginal change in our metallicity measurement during the second iteration. However, we find that this iterative process leads to more robust estimates of $\sigma_{\rm v}$ that are closer to our prior expectation.

As discussed in Section \ref{subsec:m_bin}, we do not include $\rm \lambda>1800$ \r{A} in our fitting procedure since we require the contribution of a minimum of 25 galaxies to every pixel of the composite spectrum in each stellar mass bin. Therefore, we do not use the 1935-2020 \r{A} region known as the 1978 index \citep{rix2004}, which is mainly sensitive to the iron abundance (Fe {\sc iii} $\lambda\lambda1949-1966$). We investigate the effect of excluding these regions, based on the composite spectrum of the entire sample, where there are $\sim 10\times$ more galaxies than in the stellar mass bins, allowing us to perform spectral fitting over a wider range of wavelength, $\rm \lambda=1221-2020$ \r{A}. We find that excluding the $\rm \lambda= 1800-2020$ \r{A} region from the full spectral fitting has a minimal effect on the inferred parameters. Full spectral fitting in the range of $\rm \lambda=1221-2020$ \r{A} results in slightly higher $Z_*$ value, by $\sim 0.02$ dex.

\section{Measurements of Stellar Metallicity}
\label{sec:Results}

We now use the framework built in Sections \ref{sec:smodel} and \ref{sec:fitting} to analyze composite LATIS spectra and infer their metallicities. We first consider the composite spectrum of the entire sample, and then turn to the composite spectra in stellar mass bins.

\subsection{Average metallicity of $\sim10^{10} M_\odot$ star-forming galaxies at $z\sim 2.5$}
\label{sec:Z_ave}

\begin{figure*}[!t]
    \centering
    \includegraphics[width=\linewidth]{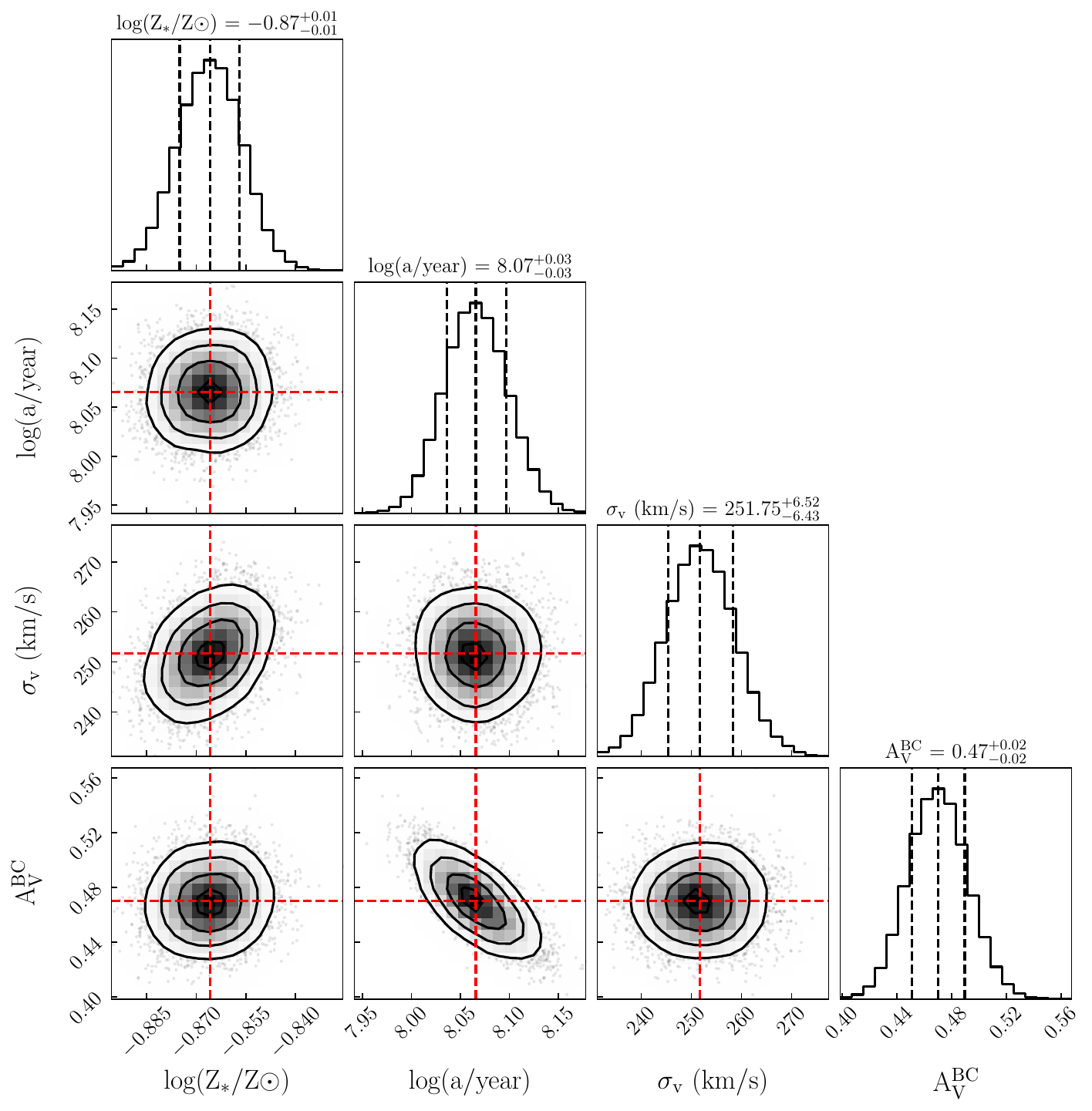}
    \caption{Corner plot showing the posterior distribution for fitted parameters on the composite spectrum of the entire LATIS sample. Contours correspond to 1$\sigma$, 2$\sigma$, 3$\sigma$ and 4$\sigma$ levels. The medians of the marginalized posteriors are indicated by red dashed lines. The black dashed lines on marginalized posteriors indicate the median
values along with the 16$^{\rm th}$ and 84$^{\rm th}$ percentiles.}
    \label{fig:corner_tot}
\end{figure*}

\begin{figure*}[!t]
    \centering
    \includegraphics[width=\linewidth]{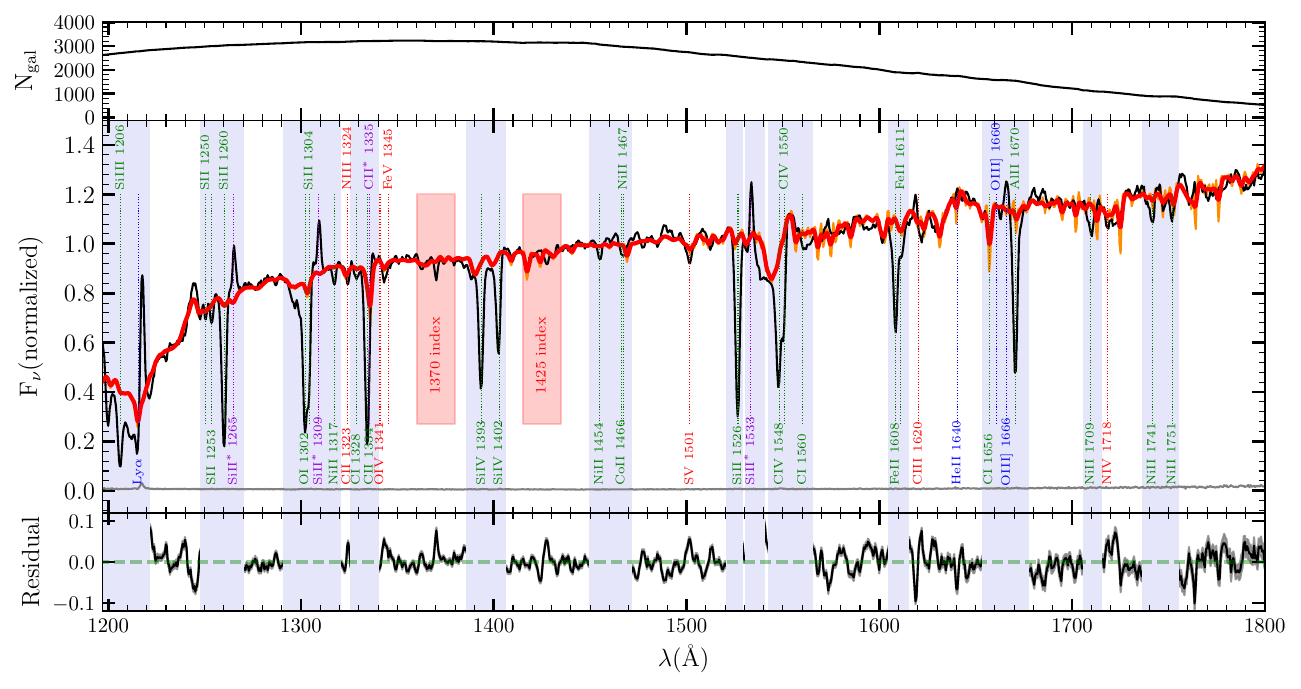}
    \caption{Similar to Figure \ref{fig:stacked_spectra_full} but the red line represents the best-fit model, while the orange line shows the unsmoothed BPASS model with its original resolution. The shaded regions in light blue, affected by interstellar absorption or nebular emission lines, are excluded from the fitting process. The bottom panel shows the fit residuals.}
    \label{fig:fit_tot}
\end{figure*}

The signal-to-noise ratio of a composite spectrum can be maximized by stacking spectra of the whole sample, resulting in lower measurement uncertainties for the sample average metallicity. We fit the BPASS models to the composite spectrum of the full sample, which has a median stellar mass of $M_*\sim10^{9.8}M_\odot$, and find that the average $\log(Z_*/Z_\odot)$ is $-0.87^{+0.01}_{-0.01}$. Figure \ref{fig:corner_tot} shows a corner plot demonstrating that all parameters are well constrained. Figure \ref{fig:fit_tot} shows that the models generally provide a good fit to the composite spectrum with root mean square (RMS) of residuals $\sim 2$\%. The RMS of the fractional noise for the composite spectrum is $\sim 1\%$. Given the very high signal-to-noise ratio of the full sample's composite spectrum, the RMS of the residuals is notably increased by model inaccuracies. Therefore, the error estimates of parameters derived by fitting the full composite spectrum may be underestimated, since they may not fully account for the model's limitations. Underestimation of errors is less of a concern for composite spectra in the stellar mass bins, where observational noise outweighs model inaccuracies.  

The stellar photospheric absorption features in the  rest-frame FUV spectrum are predominantly caused by transitions of highly ionized iron \citep[e.g.,][]{bran1998}. Therefore $Z_*$ can be translated to $\rm [Fe/H]$ by $\rm [Fe/H]\approx \log(Z_*/Z_\odot)=-0.87^{+0.01}_{-0.01}$. In this study, we measure the FUV-weighted stellar metallicity, which is expected to closely resemble the gas-phase metallicity measured in the ISM where recent star formation has occurred. In general, the FUV-weighted metallicity is expected to be marginally lower than the gas-phase metallicity by only $\lesssim 0.1$ dex \citep{kash2022}. However, similar galaxies at this redshift have an average gas-phase oxygen abundance of $\rm [O/H]\sim -0.3^{+0.1}_{-0.1}$, as derived from rest-frame optical strong lines \citep[e.g.,][]{Erb2006,Steidel14,str2018,san2021}.

\begin{figure}
    \centering
    \includegraphics[width=\linewidth]{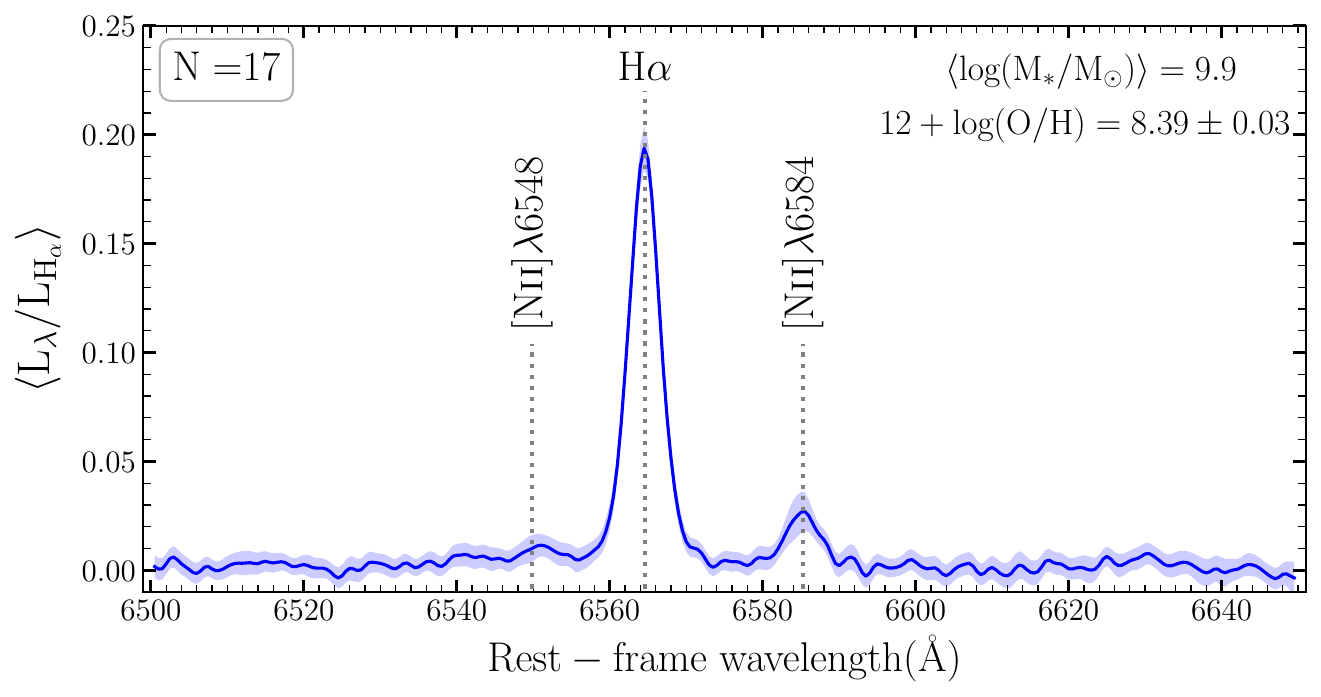}
    \caption{Composite spectrum of 17 LATIS galaxies in the rest-frame optical, obtained from Keck/MOSFIRE observations of the MOSDEF survey. Shaded regions indicate the corresponding errors obtained from bootstrapping.}
    \label{fig:latis_mosdef}
\end{figure}

To assess the gas-phase oxygen abundance within our sample, we measure $\rm [O/H]$ for the composite rest-frame optical spectra of 17 LATIS galaxies that were observed using Keck/MOSFIRE as part of the MOSDEF survey\footnote{https://mosdef.astro.berkeley.edu/for-scientists/data-releases/} \citep{Kriek15}. We ensure that these galaxies have $\rm H\alpha$ line detection with $\rm S/N\geq3$ and are not flagged as AGNs based on X-ray emission or IRAC colors. Additionally, we require that $\rm log([N${\sc ii}]$\rm \lambda 6584/H\alpha)<-0.3$ to exclude optical AGNs \citep[e.g.,][]{coi15}. The selected 17 galaxies have a median redshift of $z\sim 2.4$ with a median stellar mass of $\rm 10^{9.9}M_\odot$, which is comparable to the median mass of the entire sample. \add{Additionally, the median of $\rm log(SFR[M_\odot/yr])$ for these 17 galaxies is 1.31, which is slightly lower than the 1.54 observed for the full sample. According to \cite{san2018}, this difference in SFR corresponds to a deviation of $\sim 0.03$ dex in metallicity, which is within our measurement error, thereby ensuring the 17 galaxies are representative for purposes of metallicity comparison.} We construct the composite spectrum and measure the flux of the emission lines following the method described in \cite{chartab2021}. The resulting composite spectrum is shown in Figure \ref{fig:latis_mosdef}. We determine $\rm \langle [N${\sc ii}$]\lambda 6584/{\rm H{\alpha}}\rangle=0.13\pm0.02$, which corresponds to a gas-phase oxygen abundance of $\rm 12+\log(O/H)=8.39\pm 0.03$ ($\rm [O/H]=-0.30\pm 0.03$) using the calibrations of \cite{bia2018}. This calibration relies on direct-method metallicities obtained from stacked spectra of local analogs of $z\sim 2$ galaxies; as a result, there is no evaluation of its inherent scatter. We assume that the scatter is the same as \cite{pet04} calibration, which is 0.18 dex. Since the composite spectrum is composed of 17 galaxies, the intrinsic error in the oxygen abundance of the composite spectrum reduces to $0.18/\sqrt{17}=0.04$ \citep[e.g.,][]{Erb2006,san2015}. We add this in quadrature to the measurement uncertainty, yielding a final result of $\rm [O/H]=-0.30\pm 0.05$. We emphasize that this error is purely statistical; the dominant systematic errors are discussed below.

Oxygen is the most abundant among the $\alpha$ elements. Therefore, a measurement of $\rm [O/Fe]$ can be taken as an approximation of $\rm [\alpha/Fe]$. In this study, we find $\rm [Fe/H]=-0.87\pm 0.01$ and $\rm [O/H]=-0.30\pm 0.05$, resulting in $\rm [\alpha/Fe] \sim [O/Fe]=0.57\pm 0.05$ (statistical errors only). This result is consistent with the average $\alpha$-enhancement reported for star-forming galaxies at $z\sim 2.5$ \citep[e.g.,][]{ste2016,str2018,cul2019,kash2022}, suggesting that young high-redshift galaxies have not yet undergone significant iron enrichment through SN Ia.

There are substantial systematic uncertainties associated with gas-phase metallicity measurements. For instance, there is a systematic offset of $\sim 0.3$ dex between ``direct'' electron-temperature-based and photoionization-model-based gas-phase metallicities \citep[e.g.,][]{kew2008,blan2019}. Furthermore, there is a 0.24 dex offset between gas-phase metallicities derived from collisionally excited lines and recombination lines \citep{ste2016} in local H {\sc ii} regions. These discrepancies are of particular importance when comparing gas-phase and stellar metallicity on an absolute scale. Consequently, the errors we have estimated for $\rm [O/Fe]$ represent measurement errors and do not encompass systematic errors. Taking into account the tendency of direct method gas-phase abundances to underestimate $\rm [O/H]$ \citep[e.g.,][]{cam2023}{}{}, it becomes plausible that the $\alpha$-enhancement could be larger than our estimate.

\label{stellar_MZR}
\begin{figure*}
    \centering
    \subfloat{{\includegraphics[width=8.5cm]{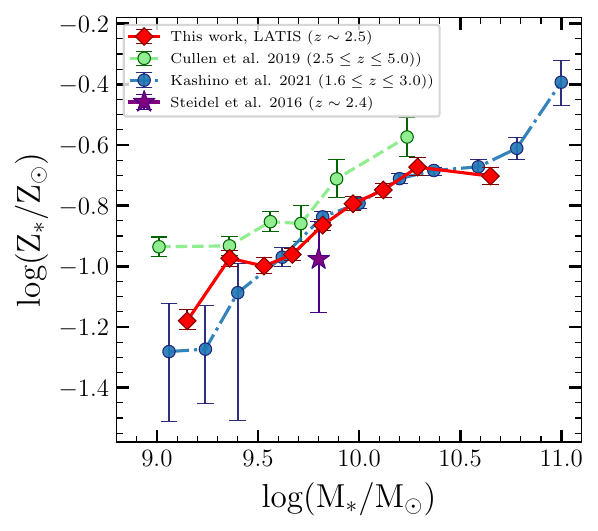} }}%
    \subfloat{{\includegraphics[width=8.5cm]{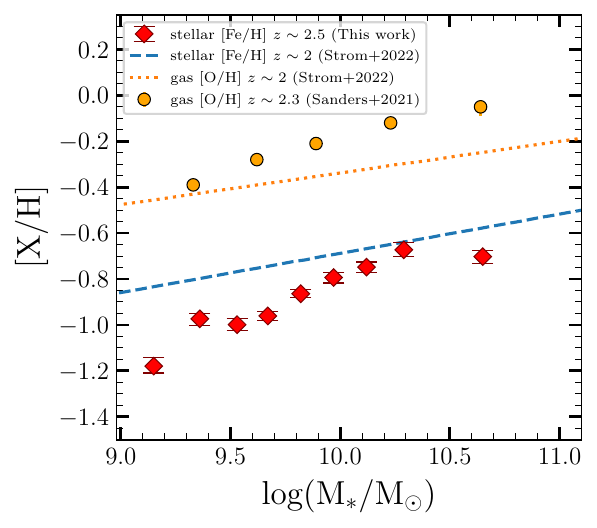} }}%

    \caption{\textit{Left:} The stellar MZR for our sample of star-forming galaxies at $z\sim 2.5$. The data points represent the posterior median metallicity derived from composite spectra in bins of stellar mass, as described in Section \ref{sec:fitting}. The error bars represent the 1$\sigma$ uncertainty. For comparison, the MZR from previous studies at similar redshifts by \cite{cul2019} and \cite{kash2022} are shown in green and blue, respectively. \textit{Right:} Comparison of our stellar $\rm [Fe/H]$ with the gas-phase $\rm [O/H]$ at $z\sim 2$ measured by \cite{san2021} and \cite{strom2022}. The blue dashed line indicates the stellar mass -- stellar $\rm [Fe/H]$ relation inferred by \citealt{strom2022} based on photoionization modeling of rest-frame nebular spectra.}
    \label{fig:MZR}
\end{figure*}

\subsection{Stellar Mass - Metallicity Relation}
\label{sec:stellarmzr}

This section presents the relationship between stellar mass and stellar metallicity for our sample at $z\sim2.5$. The results of our full spectral fitting analysis on composite spectra in bins of stellar mass, as described in Section \ref{subsec:m_bin}, are presented in Appendix \ref{app:composite_all} and summarized in Table \ref{table:composite}. As shown in the left panel of Figure \ref{fig:MZR}, the stellar $M_*-Z_*$ relation exists at $z\sim 2.5$, such that $Z_*$ increases with increasing stellar mass at low masses and flattens at $M_*\gtrsim10^{10.3}M_\odot$. This flattening has important implications for galactic winds in massive galaxies, which we will consider in Section~\ref{sec:model_eta}.

Results from other studies at relatively similar redshifts are included in the left panel of Figure \ref{fig:MZR} for comparison. Our $M_*-Z_*$ relation is in good agreement with that of zCOSMOS-deep survey from \cite{kash2022} at $1.6<z<3$. Their sample shares a comparable range of redshifts with ours and was selected in a similar manner. Additionally, they employed stellar population models similar to ours. Therefore similar results are anticipated to be obtained. However, we find slightly lower metallicities compared to the VANDELS survey \citep{cul2019}. In our analysis, we utilize the BPASS v2.2.1 models, whereas, in \cite{cul2019}, the WM-basic stellar population models were employed. They performed a comparative experiment using BPASS v2.1 models, which resulted in slightly lower metallicities by $\sim 0.09$ dex. Therefore, it is recommended to adjust their measurements by $\sim 0.1$ dex to ensure a meaningful comparison with our findings. After adjusting for this systematic offset, our result is in better agreement with \cite{cul2019}. \add{It should be highlighted that while our study focuses on $z\sim2-3$, the VANDELS survey explored a broader redshift range of $z=2.5-5$. However, \citet{cul2019} did not observe significant redshift trends within this range. The lack of such trends suggests that the different redshift ranges between our studies may not be detrimental to our metallicity comparison. Moreover, our comparison supports their observation that stellar MZR does not evolve strongly over $z\sim2-5$.}

Despite the generally satisfactory agreement with existing literature, we note that prior studies did not take into account the effect of extra attenuation in birth clouds, which may have an impact on the derived metallicity. The measurement of metallicity can be biased toward less obscured, lower metallicity environments if the extra attenuation present in the birth clouds of young stars is not taken into account \citep{Chartab2022}. To assess this effect, we repeat the analyses considering $A^{\rm BC}_{\rm V}=0$ (Appendix \ref{app:extra_dust}). We then determine the offset between the metallicity measurements and those obtained with additional dust attenuation considered for birth clouds. Our finding indicates that neglecting additional dust attenuation within birth clouds results in an underestimated stellar metallicity by $\sim 0.1$ dex.          

\begin{table*}[!ht]
\centering

\caption{Properties of composite spectra and measurements of stellar metallicity}
\label{table:composite}
\begin{tabular*}{\textwidth}{@{\extracolsep{\fill}}cccccc@{}}
\hline
\multirow{2}{*}{$\log(M_*/M_\odot)$} & \multirow{2}{*}{$\rm \langle \log{\frac{M_*}{M_\odot}}\rangle$} & \multirow{2}{*}{$\rm N_{gal}$\footnote{The number of galaxies in a stellar mass bin that is used to construct a composite spectrum.}} & median S/N & $\rm N_{min}$ & \multirow{2}{*}{$\rm log(Z_*/Z_\odot)$}\\ 
\multicolumn{1}{c}{} & & & per pixel\footnote{Median signal-to-noise ratio per pixel (100 \rm{km/s}) over a wavelength range of 1221-1800 \r{A}.} & per pixel\footnote{The minimum number of galaxies contributing per pixel in the composite spectrum at the wavelength range of 1221-1800 \r{A}.}  & \\
\hline\hline
\multicolumn{6}{c}{The entire LATIS sample} \\ \hline
9 -- 11.5 & 9.78 & 3491 & 138 & 536 & $-1.18^{+0.04}_{-0.03}$\\  \hline\hline
\multicolumn{6}{c}{Stellar mass bins} \\ \hline
9.00 -- 9.25 & 9.15 & 237 & 31 & 25 & $-1.18^{+0.04}_{-0.03}$\\
9.25 -- 9.45 & 9.36 & 448 & 45 & 30 & $-0.97^{+0.02}_{-0.03}$\\
9.45 -- 9.60 & 9.53 & 481 & 51 & 64 & $-1.00^{+0.03}_{-0.03}$\\
9.60 -- 9.75 & 9.67 & 563 & 58 & 62 & $-0.96^{+0.02}_{-0.02}$\\
9.75 -- 9.90 & 9.82 & 567 & 57 & 96 & $-0.86^{+0.02}_{-0.02}$\\
9.90 -- 10.05 & 9.97 & 434 & 49 & 92 & $-0.79^{+0.02}_{-0.02}$\\
10.05 -- 10.20 & 10.12 & 313 & 44 & 67 & $-0.75^{+0.02}_{-0.02}$\\
10.20 -- 10.40 & 10.29 & 230 & 36 & 46 & $-0.67^{+0.03}_{-0.03}$\\
10.40 -- 11.50 & 10.65 & 219 & 31 & 50 & $-0.70^{+0.03}_{-0.03}$
\\\hline\hline

\end{tabular*}
\end{table*}

In the right panel of Figure \ref{fig:MZR}, we compare our stellar MZR with the gas-phase $\rm M_*-[O/H]$ relation from \cite{san2021}. We find that there is no clear trend of $\rm [O/Fe]$ with stellar mass, which is in agreement with previous studies \citep[e.g.,][]{strom2022,kash2022}{}{}. Instead, the values remain relatively constant across all mass ranges, with an average value of $\rm [\alpha/Fe] \sim [O/Fe] \sim 0.6\pm 0.1$. This suggests that the galaxies in our sample are $\alpha$-enhanced across the entire mass range. It is also worth noting that the average $\rm [\alpha/Fe]$ value we find here is consistent with what we estimate for the composite spectrum of the entire sample in Section \ref{sec:Z_ave}.

Previous studies of the gas-phase MZR at $z\sim 2$ have not shown such a strong flattening at the massive end (right panel of Figure \ref{fig:MZR}), although \cite{san2021} reported signs of flattening for the most massive stellar mass bin at $M_*\sim 10^{10.5} M_\odot$. This difference could be partly explained by the smaller samples used in gas-phase studies, which are mostly restricted to $M_*<10^{10.5} M_\odot$, with only a handful of galaxies beyond this stellar mass. In contrast, our sample includes 219 galaxies in the last stellar mass bin of $\log(M_*/M_\odot)=10.4-11.5$, and our overall larger sample also enables finer mass bins that better define the shape of the MZR. In stellar MZR studies, \cite{cul2019} analyzed a sample of $M_*\lesssim 10^{10} M_\odot$ galaxies, the stellar mass range where we find no flattening consistent with their findings. The results of \cite{kash2022}, although not explicitly stated in their paper, suggest signs of flattening at the massive end before a sharp rise (with large uncertainty) in their last stellar mass bin at $M_* \sim 10^{11} M_\odot$. 

Moreover, the right panel of Figure \ref{fig:MZR} includes the stellar MZR at $z\sim2$ from \cite{strom2022}, who used a different method to constrain $[\rm Fe/H]$. They combined BPASS stellar population synthesis models with \Cloudy{} photoionization models to infer $[\rm Fe/H]$ from the rest-optical nebular spectra of galaxies in the Keck Baryonic Structure Survey \citep[KBSS;][]{rudie2012,Steidel14}. We find slightly lower [Fe/H] along with a more complex mass dependence than the linear form used in the \cite{strom2022} model, but the overall level agreement is still encouraging considering the very different data sets and methods underpinning these analyses.

\subsection{UV Slope Dependence of the MZR}
\begin{figure}
    \centering
    \includegraphics[width=\linewidth]{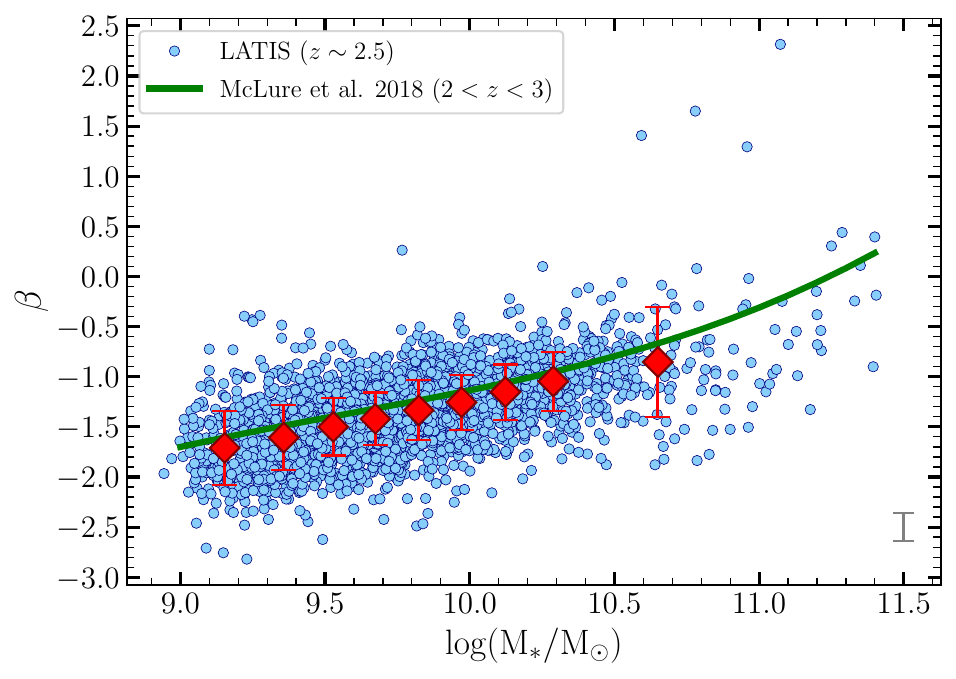}
    \caption{UV spectral slope ($\beta$) plotted against stellar mass for the LATIS sample (blue points), with red diamonds indicating the median $\beta$ in each stellar mass bin. The green line shows the $M_*-\beta$ relation from \cite{mcL2018} at $2< z < 3$ for comparison. The median uncertainty in the $\beta$ measurements is displayed in the lower right corner.}
    \label{fig:Beta-M}
\end{figure}

The flattening of the stellar MZR at high stellar masses observed in our study may have several possible explanations. One possibility is that in the highest-mass bins, our sample is potentially biased against the most metal-rich galaxies. If such galaxies are also dusty and red, they may be underrepresented in our sample due to their faint FUV fluxes, which could lead to a flattening of the MZR at high masses. Another possibility is that the physical processes governing the stellar metallicity of galaxies change at high stellar masses. In this section, we explore the impact of dust on the MZR to assess the potential biases against red, dusty galaxies at high masses. In order to determine the UV continuum slope, denoted as $\beta$, we employ a power law ($f_{\lambda} \propto \lambda^{\beta}$) fit to the photometric data (outlined in Section \ref{sec:photometry}) within the rest-frame range of $\lambda = 1300-2600$ \AA. We require our galaxies to have at least three photometric measurements in this wavelength range. We do not use spectroscopic data to measure $\beta$ since not all galaxies are covered in the desired wavelength range. Therefore, we rely on photometrically-measured $\beta$ values to ensure consistent measurement across our sample. The median uncertainty in $\beta$ is $\sim 0.14$ (Figure \ref{fig:Beta-M}). 

\begin{figure}
    \centering
    \includegraphics[width=\linewidth]{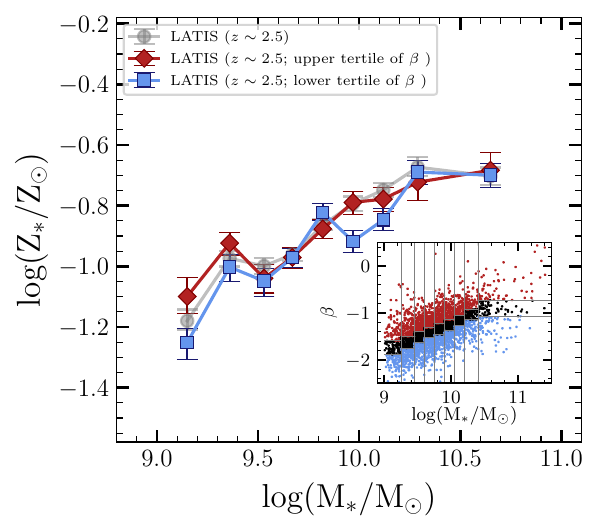}
    \caption{The stellar MZR for two subsamples of galaxies split based on the $M_* - \beta$ relation as shown in the inset.  The blue and red points represent the \add{lower and upper tertiles of $\beta$} subsamples, respectively. We find that there is no significant dependence of the MZR on $\beta$, suggesting that the flattening observed in the stellar MZR for high-mass galaxies is not heavily affected by selection effects related to color.}
    \label{fig:MZR_dust}
\end{figure}

\add{To investigate the effect of dust content on the stellar MZR, we divide the sample into three subsamples corresponding to the tertiles of $\beta$. Figure \ref{fig:MZR_dust} shows the stellar MZR for the subsamples representing the upper and lower tertiles of $\beta$}. This comparison indicates no significant dependence of the MZR on $\beta$. 
If this independence holds over the full range of $\beta$ present in a mass-limited sample, then the flattening of the MZR we observe cannot be attributed to color-related selection effects. However, we cannot guarantee that the independence of $\beta$ and metallicity extends to extremely dusty massive galaxies. By examining Figure \ref{fig:Beta-M}, we note that our sample closely follows the relation of the mass-selected star-forming sample in \cite{mcL2018} (green line) except for the very highest mass bin, where our sample is biased toward bluer galaxies. Nonetheless, at the turn-over stellar mass of the stellar MZR ($M_* \approx 10^{10.3} M_\odot$), our sample does match with the trend reported by \cite{mcL2018}. Therefore, we conclude that the observed flattening of the MZR is due to physical mechanisms rather than sample selection.

\subsection{Mass-Dependent Trends in Galactic Winds}
\label{sec:model_eta}

Analytic models provide valuable insight into the physical mechanisms governing the metallicity enrichment of galaxies. These models consider the interplay between key factors such as the inflow and outflow rates, SFR, the return of stellar material to the ISM, and changes in the total gas mass to unveil the complex mechanisms that shape the metallicity evolution of galaxies. When modeling the FUV-based stellar metallicity, which serves as a proxy for the iron abundance, it is important to incorporate the delay time distribution (DTD) of SNe Ia in order to track the evolution of iron-peak elements separately from the $\alpha$-elements. This distinction is crucial as iron-peak elements are produced by both CCSNe and SNe Ia, while $\alpha$-elements are only produced by CCSNe. Incorporating the DTD for SNe Ia in the analysis allows for a more comprehensive understanding of the metallicity evolution in galaxies.

In this work, we adopt the one-zone model of \cite{Weinberg17}, which accounts for the DTD of SNe Ia. We can then constrain the mass loading factor ($\eta=\frac{\dot{M}_{\rm outflow}}{\dot{M}_{*}}$) of $z\sim 2.5$ galaxies based on the observed stellar MZR.

The \cite{Weinberg17} model assumes that the star formation efficiency (SFE), which represents the ratio of the star formation to the gas mass available, remains constant over time (‘‘linear Schmidt law’’). The DTD for enrichment from Type Ia SNe is modeled as an exponentially declining function, with a minimum delay time of 0.15 Gyr and an e-folding timescale of 1.5 Gyr. In this work, we adopt a simple constant star formation history and the fiducial values from \cite{Weinberg17} for all free parameters of the model, except the mass loading parameter, SFE and CCSNe oxygen yield. We refer the reader to their paper for further details. 

\begin{figure}
    \centering
    \includegraphics[width=\linewidth]{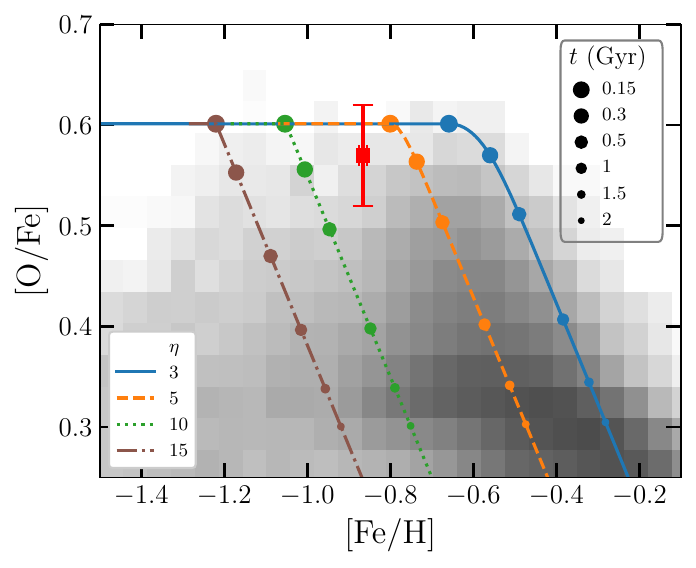}
    \caption{Chemical evolution tracks in the $\rm [O/Fe]$ -- $\rm [Fe/H]$ plane for different values of $\eta$. The different sizes of circles represent the ages of the galaxies along each track. The red square in the figure denotes our measurements for the entire LATIS sample at $z\sim 2.5$. The gray density map illustrates the distribution of APOGEE (DR17) stars without a {\tt STAR\_BAD} flag and with reliable abundance ratio values ({\tt O\_FE\_FLAG} $==$ 0 and {\tt FE\_H\_FLAG} $==$ 0). }
    \label{fig:model_alpha}
\end{figure}

The adopted values for the iron and oxygen yields in \cite{Weinberg17} result in a maximum $\rm [O/Fe]\sim0.4$ for metal enrichment entirely from CCSNe. This value is smaller than that measured in $z \sim 2.5$ galaxies (Section~\ref{sec:stellarmzr}; \citealt{ste2016}), and it is also exceed by low-metallicity stars in the Milky Way. For instance, Figure \ref{fig:model_alpha} shows that the abundances of APOGEE \citep[DR17;][]{abd2022} stars reach up to [O/Fe] $\sim 0.6$ \citep[see also][]{ste2016}, which is closer to the theoretical yield calculations of \citet{nom2006}. Therefore, we have adjusted the IMF-integrated oxygen yield to from 0.015 to 0.021 in order to achieve a plateau at $\rm [O/Fe] \approx 0.6$. It is important to highlight that this adjustment does not affect our main result, trends in the mass loading factor, which is based on iron abundances.

To estimate the SFE timescale, also referred to as the depletion timescale ($\rm M_{gas}/SFR$), we utilize the scaling relation introduced by \cite{toc2020}. They found that the depletion timescale depends mainly on the redshift and the offset from the main sequence. We compare our sample with a sample of star-forming galaxies \citep[sSFR$>10^{-10.1}$ yr$^{-1}$; ][]{Pacifici16} at $z=2-3$ in COSMOS2020 catalog and find an average offset of $\sim 0.3$ dex from the star-forming main sequence for our sample. By employing Equation 4 of \cite{toc2020}, we derive an average depletion timescale of $\sim 0.3$ Gyr for our sample at $z\sim 2.5$. 

To explore the sensitivity of the analytic model to the mass-loading parameter $\eta$, we plot tracks of $[\rm O/Fe]$, as a representative of alpha-enhancement, against $\rm [Fe/H]$ for different values of $\eta$ in Figure \ref{fig:model_alpha}.  Additionally, we include the average $\rm [Fe/H]$ and $[\rm O/Fe]$ values of the LATIS sample, represented by a red square. The different-sized circles on the $\rm [O/Fe]$ -- $\rm [Fe/H]$ tracks represent the age of a galaxy. Figure \ref{fig:model_alpha} indicates that our galaxies should be dominated by recently formed stars ($\lesssim 0.5$ Gyr) to retain such high $\rm [\alpha/Fe]$ values. This is consistent with the median age of our sample inferred from SED fitting in Section \ref{sec:SEDfitting}. 

\begin{figure}
    \centering
    \includegraphics[width=\linewidth]{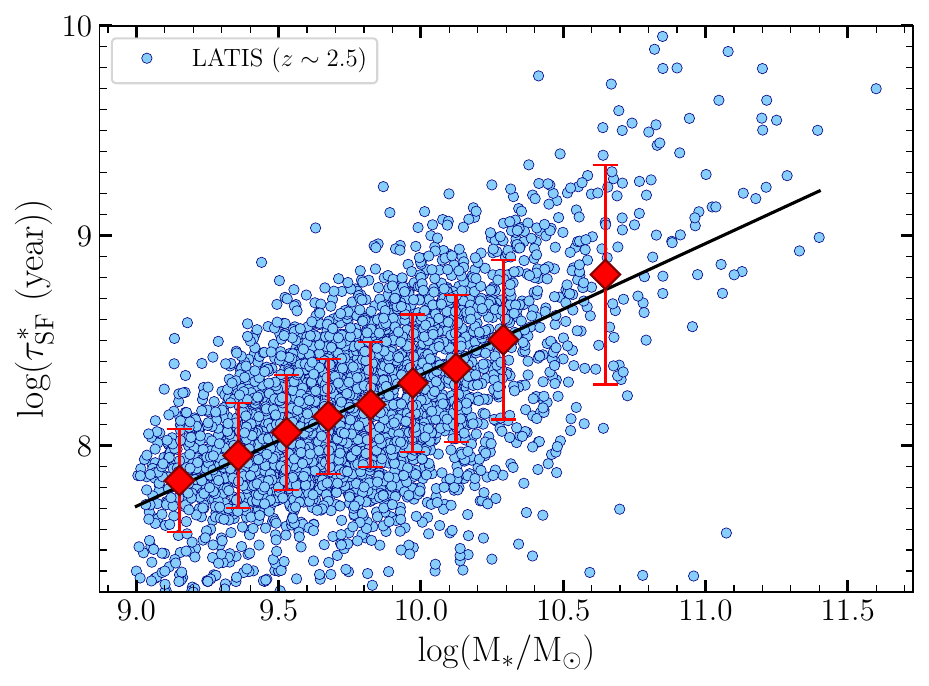}
    \caption{The characteristic star formation timescale ($\tau^{*}_{\rm SF}=M_*/$SFR) is shown as a function of stellar mass for our sample. The red diamonds represent the median values of \(\tau^{*}_{\rm SF}\) for each stellar mass bin, while the black line illustrates the best fit line to the median relationship.}
    \label{fig:age_M}
\end{figure}

\begin{figure*}
    \centering
    \includegraphics[width=0.85\linewidth]{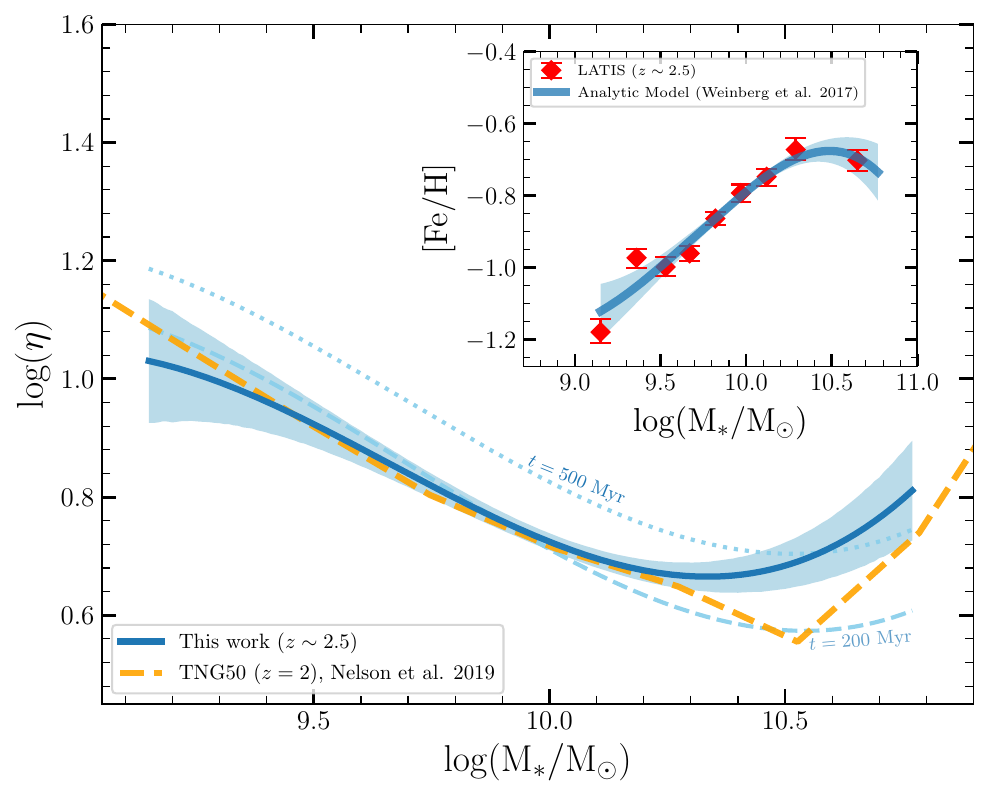}
    \caption{The mass loading factor as a function of stellar mass at $z\sim 2.5$, derived from the best-fit analytic model to our observed MZR. The relationship between the mass loading factor and the stellar mass is modeled as a cubic function. The shaded region in the main figure represents the 1$\sigma$ error around the best fit. The blue dashed and dotted lines indicate the trends for constant, mass-independent age values of 200 and 500 Myr, respectively. The orange dashed line shows the trend from TNG50 simulation \citep{nel2019} for outflows with a radial velocity greater than zero. The inset figure shows the best model MZR (blue line) along with the observed MZR (red diamonds).}
    \label{fig:model_eta}
\end{figure*}

In the following, we aim to infer the mass loading parameter $\eta$ and its dependence on stellar mass $M_*$ using the observed stellar MZR. We incorporate the $M_*$-dependent ages of galaxies into the chemical evolution modeling, estimated as the median of the characteristic star formation timescale, defined as $\tau^{*}_{\rm SF}=M_*/$SFR, at a given stellar mass (Figure \ref{fig:age_M}). Along with the other parameters described above, we can now use the \citet{Weinberg17} model to compute [Fe/H] given $\eta$ and $M_*$\footnote{We compute the model ISM metallicity at the end of the star-formation history and neglect its difference with the measured FUV-weighted stellar metallicity, which is expected to be only $\sim$0.02 dex at $z \sim 2.2$ \citep{kash2022}.}. We utilize a cubic functional form for the mass loading parameter $\eta(M_*)$ and determine the polynomial coefficients by fitting the model to the observed $\rm [Fe/H]$-$M_*$ relation. The derived mass loading parameter, presented in Figure \ref{fig:model_eta}, indicates that it declines as the stellar mass increases, but flattens and possibly even increases for $M_*\gtrsim 10^{10.5}M_\odot$. 

We note that the absolute value of $\eta$ is indeed sensitive to the various assumptions in the model, yet the overall trend remains consistent. Blue dashed and dotted lines in Figure \ref{fig:model_eta} illustrate the trend for alternative mass-independent values of age, 200 Myr and 500 Myr, respectively. In all cases, a decline in $\eta$ is observed as the stellar mass increases for low-mass galaxies; however, this trend plateaus/reverses at $M_* \sim 10^{10.5}M_\odot$.

\citet{san2021} derived the mass-dependence of $\eta$ by applying a different chemical evolution modeling framework to observations of the gas-phase MZR at $z = 2.3$ and 3.3. They found that the metal mass-loading factor $\zeta(M_*)$, which differs from $\eta$ by the ratio of the outflow and ISM metallicities, follows $\zeta(M_*) \propto M_*^{-0.35 \pm 0.02}$ at low masses, and they showed that their data were compatible both with models in which $\zeta$ flattens at high masses $M_* \gtrsim 10^{10.5} M_\odot$ and models in which $\zeta$ continues to decline. We find that $\eta \propto M_*^{-0.35 \pm 0.01}$ at low masses $M_* \lesssim 10^{10.25} M_\odot$, in excellent agreement with \cite{san2021}, but our data clearly prefer a flattening at high masses. Our estimates of $\eta$ are systematically higher than those of \cite{san2021} by $\sim 0.3$ dex, but we expect that the absolute value of $\eta$ is significantly more sensitive to modeling assumptions than the slope of $\eta(M_*)$.

To better understand the behavior of the mass loading parameter as a function of stellar mass, we compare our results with those obtained from the TNG50 cosmological simulation (orange dashed line in Figure {\ref{fig:model_eta}}). TNG50 is a cosmological hydrodynamic simulation within a 50 Mpc box sampled by $2160^3$ gas cells, resulting in a baryon mass resolution of $8 \times 10^4 M_\odot$ and an average spatial resolution $\sim 100-200$ pc in ISM gas. \cite{nel2019} measured the mass loading parameter at $z\sim2$ in TNG50 and found a similar nonmonotonic behavior as a function of galaxy stellar mass. Specifically, they found that the mass loading parameter turns over and rises rapidly above $10^{10.5} M_\odot$, consistent with our finding. In the simulation, this behavior is traced to winds driven by supermassive black holes, which dominate over stellar-driven winds in massive galaxies.

\section{Discussion}
\label{sec:Discussion}

Our results show that the stellar MZR exists at $z\sim 2.5$, such that stellar metallicity increases with increasing stellar mass at low masses. However, we also find evidence of flattening at the massive end of the MZR ($M_*\gtrsim 10^{10.3} M_\odot$).

\begin{figure}
    \centering
    \includegraphics[width=8.5cm]{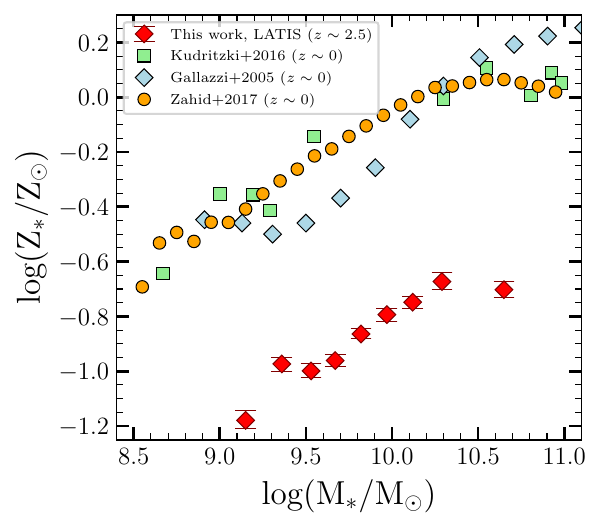}
    \caption{Comparison of our stellar MZR at $z\sim 2.5$ with those of \cite{zah2017}, \cite{kud2016} and \cite{gal2005} measured at $z\sim 0$. Our $z\sim 2.5$ MZR exhibits a significant shift toward lower metallicities compared to the $z\sim 0$ relation, while the slopes of the two relations are consistent. }
    \label{fig:MZR_z0_comp}
\end{figure}

In Figure \ref{fig:MZR_z0_comp}, we compare stellar MZR measurements at $z \sim 0$ with our result at $z\sim 2.5$. Using full spectral fitting to the rest-frame optical spectra, \cite{zah2017} measured the stellar MZR for $\sim 200000$ star-forming galaxies in the Sloan Digital Sky Survey (SDSS). Their result is in agreement with the relation derived by \cite{kud2016} based on analysis of individual supergiant stars. The results of \cite{zah2017} differ from those of \cite{gal2005}, possibly due to the fact that the latter used Lick indices to measure metallicity and included data from both star-forming and quiescent galaxies in their analysis. The slope of our stellar MZR at $z\sim 2.5$ below the turnover mass ($M_*\sim 10^{10.3} M_\odot$) is 0.40 $\pm$ 0.04, which agrees with the slope meausured at $z\sim0$ by \cite{zah2017}. However our $z\sim 2.5$ MZR shows a shift toward lower metallicities by $\sim 0.7$ dex compared to the local stellar MZR. We note that the actual offset is greater, as $z\sim 0$ metallicities in \cite{zah2017} are weighted by optical luminosity, and when converted to FUV-weighted for comparison with our measurements, they need to be shifted upwards by $\sim 0.1$ dex to account for the differences in the stellar populations being traced \citep{kash2022}. Therefore, the average stellar metallicity of galaxies at fixed stellar mass increases by a factor of $\sim5$ from $z\sim2.5$ to 0 (see also \citealt{cul2019}). However, \cite{san2021} reported that the gas-phase metallicity $\rm [O/H]$ only increased by a factor of 2 in the same time interval, leading to a discrepancy between the evolution of the FUV-weighted stellar metallicities and gas-phase oxygen abundances. While it is expected that these two metallicities evolve somewhat differently due to the different timescales that they trace, they should still closely follow each other with a difference of $\sim 0.1-0.2$ dex \citep{kash2022}. Thus, the large observed difference in their evolution is likely due to delayed iron enrichment by SN Ia.

\add{We note that our sample is primarily positioned on or above the star formation main sequence, with a median offset of $\sim 0.3$ dex. At low redshifts, SFR is anticorrelated with gas-phase metallicity at fixed stellar mass \citep[e.g.,][]{man2010,cur2020}. Although the existence and form of the ``fundamental metallicity relation'' at higher redshifts is debated, \cite{san2018} found that, at a fixed stellar mass, the relationship $\Delta \log(\rm{O/H}) \sim -0.15 \times \Delta \log(\rm{SFR})$ holds for star-forming galaxies at $z \sim 2.3$. If a similar relation applies to the iron abundance of young stars, then our estimates of the stellar metallicity would be $\sim$0.05 dex lower than that of a mass-limited sample. This is a minor potential bias and is not likely to affect our main results on the shape of the stellar MZR.}

Our MZR observations imply a transition in the characteristics of galactic winds around masses of $M_* \approx 10^{10.3}M_\odot$, but they cannot uniquely identify the physical origin of this transition. For further insights we turn to simulations and other observations. We find that our $M_*-\eta$ relation, derived using an analytic chemical evolution model, matches closely the one measured in the TNG50 simulation at $z = 2$. In the simulation, the mass loading attributed to star formation declines monotonically with stellar mass, and the upturn in $\eta$ at high masses $M_* \gtrsim 10^{10.5} M_\odot$ is caused by winds driven by supermassive black holes (SMBHs; \citealt{nel2019}). In the TNG model, low-luminosity AGN in this population drive high-velocity winds that expel cool, dense gas from the nuclear ISM. These outflows eventually lead to the ``inside-out'' quenching of star formation.

Furthermore, independent observations support the widespread occurrence of AGN-driven outflows in a similar population of $z \sim 1$-2 galaxies. \citet{gen2014} and \citet{ForsterSchreiber14} find a rapid increase at masses $M_* \gtrsim 10^{10.9} M_\odot$ of the incidence of broad, nuclear emission lines with line ratios indicative of shocks or AGN photoionization. This mass scale is higher than the mass at which we find a turnover of the MZR. However, it is possible that AGN are driving winds in slightly lower-mass galaxies that are sufficient to affect the chemical evolution but are not as readily detectable via optical emission line diagnostics. This possibility might be tested by observations using the higher angular resolution  afforded by extremely large telescopes.  Taken together, these additional lines of evidence suggest that SMBH-driven winds in massive galaxies are the most likely origin of the flattening of the stellar MZR that we observe.

\section{Summary}
\label{sec:Summary}

In this paper, we present the stellar mass - stellar metallicity relation for 3491 star-forming galaxies at $2 \lesssim z \lesssim 3$ using rest-frame FUV spectra from the LATIS survey. We utilize the BPASS (v2.2.1) synthesis models to fit high signal-to-noise composite spectra of galaxies. Our findings can be summarized as follows:

\begin{itemize}
    \item We find that the stellar metallicity increases monotonically with increasing stellar mass at lower masses but flattens at $M_*\gtrsim 10^{10.3}M_\odot$. The slope of our $z\sim 2.5$ MZR at the low mass end is similar to that of local stellar MZR. However, there is a significant offset of $\sim 0.7$ dex toward lower metallicities compared to the local stellar MZR.   

    \item We compare our measurements of stellar metallicity to the gas-phase oxygen abundance of galaxies at similar redshift. We find no clear trend in the $\rm[O/Fe]$ - $M_*$ relation. $\rm[O/Fe]$ remains relatively constant across the full mass rangewith an average value of $\rm [O/Fe] \sim 0.6$, suggesting that young galaxies at $z \sim 2.5$ have yet to experience substantial iron enrichment resulting from SN Ia.

    \item Using an analytical model of chemical evolution, we constrain the mass loading parameter as a function of stellar mass. We find that the mass loading parameter decreases with stellar mass at low masses but plateaus or reverses at $M_* \sim 10^{10.5}M_\odot$. In combination with the TNG simulations and other observations, our results suggest that SMBH-driven outflows in massive galaxies at $z\sim 2.5$ are responsible for the upturn in the $M_*-\eta$ relation, which in turn explains the flattening of the MZR at the massive end. Massive galaxies undergo strong SMBH-driven outflows, which remove the metal-rich gas from the ISM.

\end{itemize}

\begin{acknowledgments}

\add{We thank the anonymous referee for providing insightful comments and suggestions that improved the quality of this work.} This paper includes data gathered with the 6.5 meter Magellan Telescopes located at Las Campanas Observatory, Chile. We thank the staff at Las Campanas Observatory for their dedication and support. N.C. and A.B.N. acknowledge support from the National Science Foundation under Grant No. 2108014. \add{G.A.B. acknowledges the support from the ANID Basal project FB210003.}  
\end{acknowledgments}

\software{NumPy \citep{harris2020array}, Matplotlib \citep{matplotlib}, Astropy \citep{astropy2013,astropy2018,astropy2022}, 
PyMultiNest \citep{Buc2014}.  }

\bibliography{LATIS_MZR}

\appendix
\counterwithin{figure}{section}

\section{Best-fit spectra in stellar mass bins}
\label{app:composite_all}
\begin{figure*}[h]
    \centering
    \includegraphics[width=0.89\textwidth]{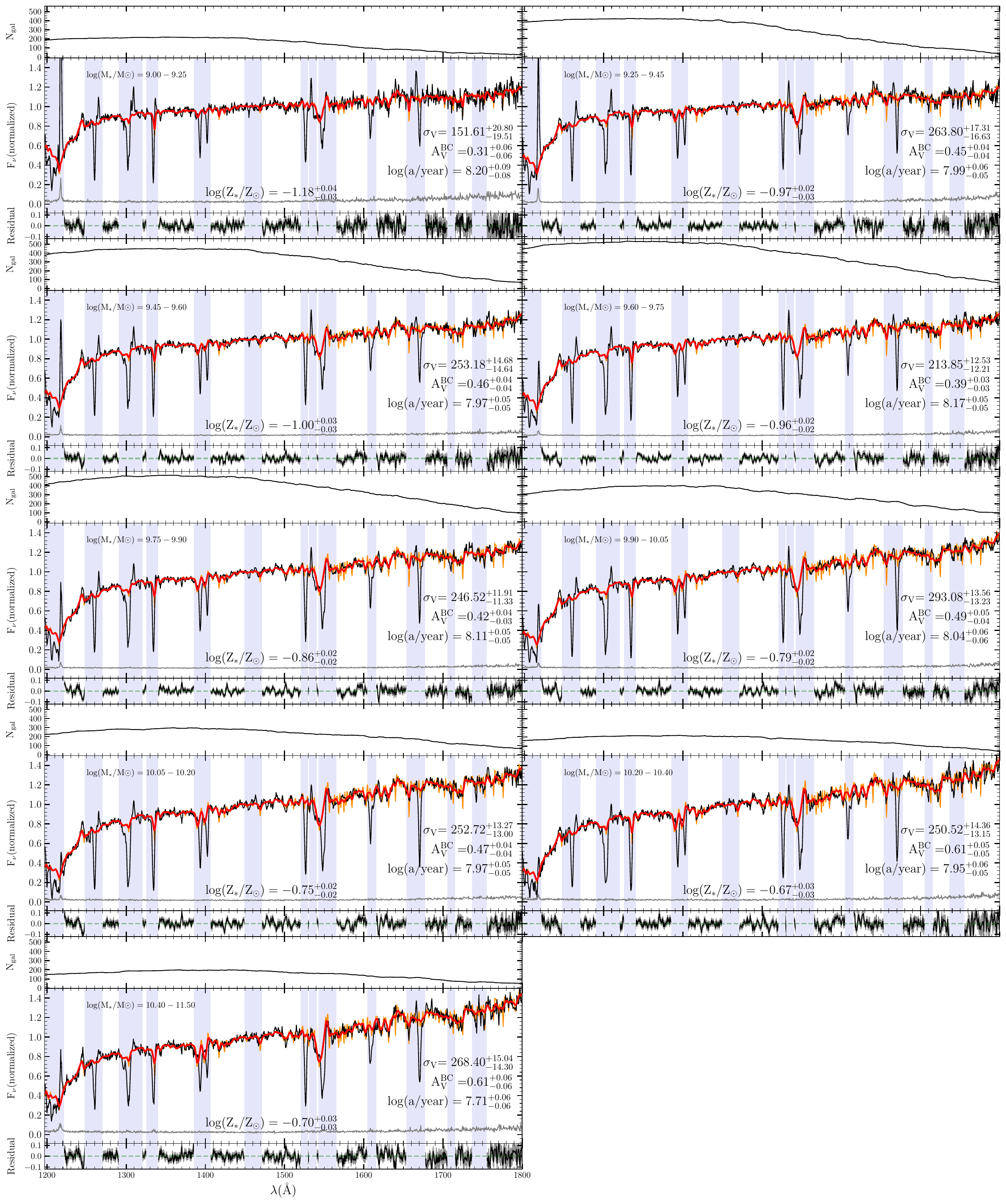}
    \caption{Similar to Figure \ref{fig:fit_tot} but for each stellar mass bin. The best-fitting parameters are indicated in legends.}
    \label{fig:fit_all}
\end{figure*}

In this section, we present fits to the LATIS composite spectra in bins of stellar mass. Each panel in Figure \ref{fig:fit_all} illustrates one of the nine composite spectra described in Section \ref{sec:stack_spectra}, along with the best-fitting parameters.

\section{Effect of $\rm A_{V}^{BC}$ on the metallicity measurements}
\label{app:extra_dust}
\begin{figure*}[h]
    \centering
    \includegraphics[width=\textwidth]{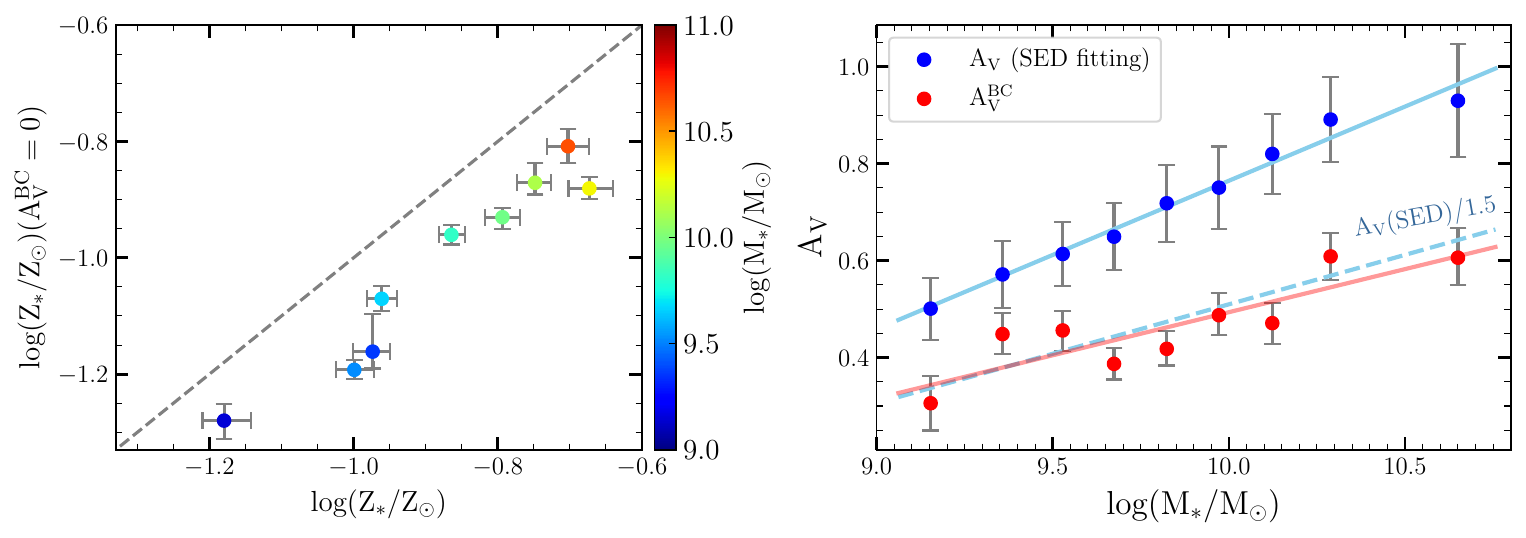}
    \caption{\textit{Left:} Comparison between the $Z_*$ measurements obtained with and without considering extra attenuation due to birth clouds ($\rm A_V^{BC}$). The x-axis represents the derived values with extra attenuation from birth clouds, while the y-axis represents the values obtained with $\rm A_V^{BC}=0$. The gray dashed line shows the one-to-one relation. \textit{Right:} The average visual band attenuation ($\rm A_V$) obtained from SED fitting (shown in blue) and $\rm A_V^{BC}$ (shown in red) as a function of stellar mass. The solid cyan and red lines show the best fit lines for the $\rm M_*-A_V$ and $\rm M_*-A_V^{BC}$ relations, respectively. The dashed cyan line represents the trend corresponding to two-thirds of the dust attenuation obtained from SED fitting.}
    \label{fig:A_v_BC_comp}
\end{figure*}

In this section, we re-measure the $Z_*$ values under the assumption that there is no additional attenuation caused by birth clouds ($\rm A_V^{BC}=0$). The left panel in Figure \ref{fig:A_v_BC_comp} illustrates a comparison between the values obtained by considering extra attenuation in birth clouds (x-axis) and those obtained with $\rm A_V^{BC}=0$ (y-axis). We find that disregarding $\rm A_V^{BC}$ leads to an underestimation of stellar metallicity by $\sim 0.1$ dex. In the right panel of Figure \ref{fig:A_v_BC_comp}, we compare the measurements of $\rm A_V^{BC}$ with the average visual band attenuation ($\rm A_V$) obtained from SED fitting, as described in Section \ref{sec:SEDfitting}, and find that $\rm A_V^{BC}/A_V\sim 2/3$. This implies a nebular to stellar attenuation ratio of $\rm (A_V+A_V^{BC})/A_V \approx 1.7$, which is  relatively consistent with previous studies in the literature \citep[e.g.,][]{pri2014,the2019,shiv2020}.

\section{The Stellar MZR with a Constant $\sigma_{\rm v}$}
\label{app:mzr_const_sigma}
\twocolumngrid

\begin{figure}
    \centering
    \includegraphics[width=0.43\textwidth]{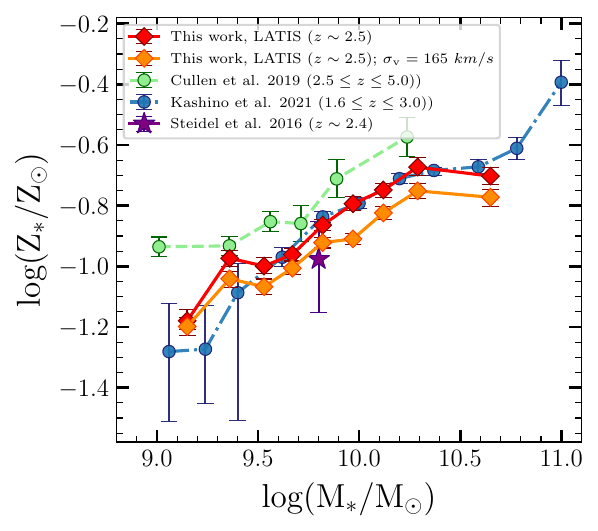}
    \caption{Similar to the left panel of Figure \ref{fig:MZR}, but with an orange curve representing the MZR when $\sigma_{\rm v}$ is fixed.}
    \label{fig:sigma_const}
\end{figure}

In this appendix, we repeat our analysis with a constant $\sigma_{\rm v}$ value to investigate its effect on the stellar MZR.  The metallicities are measured using the same methodology as described in Section \ref{sec:fitting}, except that we fix $\sigma_{\rm v}$ at the expected value of 165 km/s. The resulting MZR is presented in Figure \ref{fig:sigma_const}. We find that fixing $\sigma_{\rm v}$ to 165 km/s leads to slight lower estimate of metallicity by $\sim 0.07$ dex. However, it is noteworthy that the overall shape of the MZR remains unchanged, including the characteristic flattening at the high-mass end. Thus, we conclude that despite the sensitivity of absolute metallicities to the effective spectral resolution $\sigma_{\rm v}$, the fundamental shape of the MZR is robust and unaffected.

\end{document}